\documentclass[useAMS]{mn2e}

\usepackage{graphicx}
\usepackage{color}

\title[C, N, O, and Na abundances of RR Lyr stars]
{Surface C, N, O, and Na abundances of RR Lyrae variables
implying the nature of internal mixing in low-mass stars
}

\author[Y. Takeda]
{Yoichi Takeda\thanks{E-mail:
ytakeda@js2.so-net.ne.jp}\footnotemark[0] 
\\
11-2 Enomachi, Naka-ku, Hiroshima-shi 730-0851, Japan\\
}
\begin{document}

\date{Accepted 2022 May 18. Received 2022 April 26; in original form 2022 March 16}


\maketitle

\label{firstpage}

\begin{abstract}
Photospheric abundances of C, N, O, and Na were determined by applying the synthetic 
spectrum-fitting technique to 34 snap-shot high-dispersion spectra of 22 RR~Lyr 
stars covering a metallicity range of $-1.8 \la$~[Fe/H]~$\la 0.0$, with an aim of
investigating the mixing mechanism in the interior of low-mass giant stars by examining 
the abundance anomalies of these elements possibly affected by the evolution-induced 
dredge-up of nuclear burning products. 
Special attention was paid to check the recent theoretical stellar evolution simulations 
indicating the importance of thermohaline mixing in low-mass stars ($M \la$~1~M$_{\odot}$), 
which is expected to be more significant as the metallicity is lowered.
By inspecting the resulting abundances in comparison with those of unevolved 
metal-poor dwarfs at the same metallicity, the deficiency in C as well as enrichment 
in N was confirmed (while O is almost unchanged), the extent of peculiarities tending 
to increase with a decrease in [Fe/H]. Accordingly, the [C/N] ratio turned out to 
progressively decrease towards lower metallicity from $\sim 0$ ([Fe/H]~$\sim 0$) to
$\sim -1$ ([Fe/H]~$\sim -1.5$), which is reasonably consistent with the theoretical 
prediction in the presence of thermohaline mixing.  
However, these RR Lyr stars do not show any apparent Na anomaly (i.e., essentially 
the same [Na/Fe] vs. [Fe/H] trends as those of dwarfs), despite that metallicity-dependent 
overabundance in Na is theoretically expected for the case of non-canonical mixing. 
This inconsistency between C/N and Na may suggest a necessity of further improvement 
in the current theory.
\end{abstract}


\section{Introduction}

After a star in the core hydrogen-burning phase (main sequence) has exhausted its 
central fuel, it begins to move toward the red giant stage in the Hertzsprung--Russell 
(HR) diagram while decreasing the surface temperature ($T_{\rm eff}$) as well as 
increasing the luminosity ($L$). Because of the deepening of convention zone caused 
by the reduction of $T_{\rm eff}$, the H-burning product in the interior can be
salvaged into the surface by this enhanced mixing, 
by which the surface chemical composition of red giants tends to show 
characteristic anomalies due to the contamination of nuclear-processed material.
For example, an appreciable C-deficiency as well as N-enrichment (while O is much less 
susceptible) would result by mixing of CNO-cycle products and a moderate overabundance 
of Na may arise due to NeNa-cycle products. Likewise, $^{12}$C/$^{13}$C isotope 
ratio as well as Li or Be abundances may also serve as useful indicators.  
Yet, such an evolution-induced enhancement of convective dredge-up (standard mixing)
is not the only mechanism that would take place in the stellar interior.
It has been often argued (see, e.g., Lagarde et al. 2012b and the references therein) 
that other non-canonical mixing processes may be significant depending on situations; 
such as rotation-induced mixing (meridional circulation or shear instability) and 
thermohaline mixing (instability induced by molecular weight inversion created by 
$^{3}$He~+~$^{3}$He~$\rightarrow$~2p~+~$^{4}$He 
reaction in the external wing of the H-burning shell).

Especially, the importance of thermohaline mixing on the surface composition is theoretically 
expected for the case of low-mass ($M \la 1 \,{\rm M}_{\odot}$) metal-poor giants, 
because this process becomes more efficient as the initial mass decreases and the metallicity
is lowered. This is demonstrated in Fig.~6 or Fig.~8 of Lagarde et al. (2019), where 
the simulated [C/N] (C-to-N abundance ratio) is plotted against [Fe/H] (metallicity).\footnote{
As usual, [X/H] is the differential abundance for element X of a star relative to the Sun 
defined as [X/H] $\equiv$ $A_{*}$(X) $-$ $A_{\odot}$(X), where $A$(X) is the 
logarithmic number abundance of element X (so normalised with respect to H as $A$(H) = 12.00).  
Likewise, the notation [X/Y] is defined as [X/Y] $\equiv$ [X/H] $-$ [Y/H].}
It is evident from their figures that the [C/N] vs. [Fe/H] relations computed for 
the two cases of [1] standard mixing and [2] non-canonical mixing 
(including thermohaline mixing)\footnote{
Regarding the term ``non-canonical mixing'', three kinds of mixing processes 
are generally involved (cf. Lagarde et al. 2012b):
(i) convective (standard) mixing, (ii) thermohaline mixing, and (iii) rotational mixing.
Among these, the last rotational one may be regarded as 
insignificant in low-mass ($M \la$~1~M$_{\odot}$) stars of our concern, because efficient
rotational braking (loss of angular momentum) would take place when they are on the main 
sequence (late-type dwarfs). Accordingly, when non-canonical mixing is referred to 
in the remainder of this paper, we do not explicitly mention about rotational mixing but 
confine only to thermohaline mixing (e.g., the term ``thermohaline mixing'' is occasionally 
used to mean the equivalent concept of ``non-canonical mixing'' for simplicity).
In Lagarde et al.'s (2019) population-synthesis calculations for [C/N] where comparatively 
lower-mass stars are primarily concerned, the effect of rotation is not included 
in the non-standard mixing (regarding the impact of rotational mixing, see the 
other papers of the same group; e.g., Charbonnel \& Lagarde 2010; Lagarde et al. 2012a).   
} 
are markedly different from each other; that is,
while $|$[C/N]$|$ (the extent of mostly negative [C/N]) tends to be gradually decrease 
with a decrease in [Fe/H] (i.e., mixing is suppressed as metallicity is lowered) 
in the former canonical case, a sequence of increasing $|$[C/N]$|$ towards lower [Fe/H]
([C/N] decreases from $\sim -0.5$ to $\sim -1$ with a reduction of [Fe/H] from $\sim 0$ 
down to $\sim -1$; which means an enhanced mixing with decreasing metallicity) 
newly appears in the latter non-canonical case.

Actually, Lagarde et al. (2019) showed in their Fig.~8 that observed [C/N] values of red giants 
in several metal-poor open clusters (down to [Fe/H]~$\sim -1$) appear to be more consistent 
with the latter trend than the former one, which suggests that thermohaline mixing may play
the dominant role (over the standard mixing) in the metal-poor regime.
However, since red giant stars in a cluster have various masses and evolutionary stages
(but with almost the same age and metallicity), they are not necessarily adequate to 
study how the efficiency of mixing depends upon metallicity. 
As a matter of fact, the observed [C/N] values within the same cluster plotted in 
Lagarde et al.'s (2019) Fig.~8 are considerably diversified (presumably due to the 
differences in stellar parameters). Although Lagarde et al.'s (2019) theoretical 
population-synthesis calculations for the expected surface [C/N] ratios take into 
account such parameter differences of individual stars, it is rather difficult to 
compare these theoretical [C/N] vs. [Fe/H] distributions with those derived from 
stars of open clusters, which have large dispersions and are available at only 
several ``discrete'' [Fe/H] values.

In this context, it may be worth paying attention to RR~Lyr stars, which belong to the 
class of horizontal-branch (HB) stars (post red-giant stage after the He-ignition)
currently crossing the Cepheid instability strip.  These pulsating variables may serve as 
an adequate testing bench for investigating the metallicity-dependence of mixing based on 
their surface compositions for the following reasons. (i) They are low-mass stars 
($M \la 1 \,{\rm M}_{\odot}$; typically around $\sim 0.8\,{\rm M}_{\odot}$) and have effective 
temperatures in a rather narrow range ($T_{\rm eff} \sim$~6000--7500~K), which means that 
they are similar in terms of stellar parameters and evolutionary stages. (ii) Meanwhile, 
their metallicities distribute in a sufficiently wide range ($-2 \la$~[Fe/H]~$\la 0$), 
which makes them suitable for studying the metallicity effect.  

Very recently, an interesting report following this line has been made by Andrievsky et al. 
(2021). They determined the C, N, and O abundances of nine field RR Lyr stars, and found 
that the resulting [C/N] ratios tend to decrease towards lower metallicity from [Fe/H]~$\sim 0$ 
down to $\sim -2$ (cf. Fig.~8 therein). This tendency is almost consistent with the simulation 
of Lagarde et al. (2019), supporting the importance of thermohaline mixing in low-mass giants. 
However, since Andrievsky et al.'s (2021) analysis is limited to only 9 stars, their conclusion 
needs to be confirmed based on a larger sample of RR Lyr stars.

Conveniently, useful spectral data of some 20 RR~Lyr stars applicable to this purpose 
are available, which were observed by the author's group $\sim 20$ years ago.
A part of these data for the specific 4 stars (for each of which several observations at 
different times were made) were first analysed by Takeda et al. (2006; hereinafter 
referred to as Paper~I) in order to check the consistency of O and Fe abundances 
(and the reliability of spectroscopically established atmospheric parameters) 
determined from the spectra of different pulsation phases.
Successively, based on all the available data, Liu et al. (2013; hereinafter Paper~II) 
determined the abundances of 15 elements (Na, Mg, Al, Si, Ca, Sc, Ti, V, Cr, Mn, Fe, Ni, Cu, Y, 
and Ba) for 23 RR~Lyr stars, though light elements such as CNO were outside the scope of their study. 

Accordingly, in order to provide an independent follow-up study after Lagarde et al. (2019)
and Andrievsky et al. (2021), it was decided to newly determine the photospheric abundances 
of four key elements (C, N, O, and Na) for this sample of RR~Lyr variables 
(covering the metallicity range between [Fe/H]~$\sim -2$ and $\sim 0$), 
where the main attention is paid to clarifying the trend of [C/N] 
in terms of metallicity. This is the purpose of this investigation.      

\section{Programme stars and observational data}

The observational materials used in this study are the 34 high-dispersion spectra of
22 RR~Lyr stars (repeated observations were done for 4 stars; DH~Peg, DX~Del,
RR~Lyr, and VY~Ser), which were obtained on 2004 June 27 (UT) by using the High-Dispersion 
Spectrograph (HDS) on the Nasmyth platform of the 8.2~m Subaru Telescope atop Mauna Kea. 
See Sect.~2 of Paper~I for more details on the observations. The list of these 22 programme 
stars along with their observational times (Julian day) is presented in Table~1. 
Besides, additional information (e.g., pulsation period, phase, radial velocity, etc.) 
is given in Table~1 of Paper~II, which describes essentially the same data as adopted 
in this paper (except that two spectra of V445~Oph analysed in Paper~II were discarded 
in this investigation as remarked in Sect.~3).

As to the comparison stars, Sun (G2~V star) and Procyon (F5~IV--V star of near-solar 
composition) were chosen, for which Kurucz et al.'s (1984) solar flux atlas and the 
spectra published by Takeda et al. (2005a) as well as  Allende Prieto et al. (2004)
were used as the observational data, respectively, while their atmospheric parameters 
(given in Table~1) are the same as adopted in Takeda et al. (2005b).

\setcounter{table}{0}
\begin{table*}
\begin{minipage}{180mm}
\small
\caption{Atmospheric parameters of the program stars and the resulting abundance ratios.}
\begin{center}
\begin{tabular}{cccccccccccl}\hline
\hline
Star & HJD & $T_{\rm eff}$ & $\log g$ &  $v_{\rm t}$  & [Fe/H] & [C/Fe] & [N/Fe] & [O/Fe] & [Na/Fe] & [C/N] & Remark\\ 
  (1) & (2) &     (3)       &   (4)  & (5) &  (6) &  (7) &  (8) &  (9) &  (10)  & (11) & (12) \\
\hline
AA Aql   &  2453184.955 &  6550 &  2.87 &  2.8 & $-$0.36 & $-$0.15 & +0.30 & +0.39 & +0.07 & $-$0.45  & \\
AO Peg(1)&  2453185.031 &  6180 &  2.19 &  3.1 & $-$1.23 & $-$0.47 &($-$0.09)& +0.62 & $-$0.55 &($-$0.38) & \\
AO Peg(2)&  2453185.039 &  6290 &  2.67 &  3.3 & $-$1.16 & $\cdots$ & $\cdots$ & +0.77 & $-$0.54 & $\cdots$  & \\
BR Aqr   &  2453185.011 &  6440 &  2.42 &  2.8 & $-$0.78 & $-$0.24 & +0.46 & +0.53 & +0.03 & $-$0.70  & \\
CI And   &  2453185.094 &  6380 &  2.72 &  2.8 & $-$0.39 & $-$0.14 & $-$0.04 & +0.19 & $-$0.23 & $-$0.10  & \\
CN Lyr   &  2453184.825 &  6380 &  2.98 &  3.2 & $-$0.08 & $-$0.16 & +0.07 & +0.04 & $-$0.12 & $-$0.23  & \\
DH Peg(1)&  2453184.962 &  7780 &  2.85 &  1.7 & $-$1.05 & $-$0.36 & +0.71 & +0.46 & $-$0.03 & $-$1.06  & Paper~I \\
DH Peg(2)&  2453185.048 &  6990 &  2.69 &  1.7 & $-$1.21 & $-$0.23 & +0.72 & +0.56 & +0.02 & $-$0.96  & Paper~I \\
DH Peg(3)&  2453185.121 &  7110 &  2.79 &  1.7 & $-$1.23 & $-$0.26 & +0.85 & +0.60 & +0.07 & $-$1.11  & Paper~I \\
DM Cyg   &  2453184.988 &  6400 &  2.88 &  3.1 & +0.02 & $-$0.14 & $-$0.13 & +0.02 & $-$0.10 & $-$0.01  & \\
DO Vir   &  2453184.779 &  6340 &  2.63 &  2.0 & $-$1.24 & +0.07 & +0.78 & +0.49 & +0.22 & $-$0.71  & \\
DX Del(1)&  2453184.893 &  6670 &  2.63 &  3.1 & $-$0.21 & $-$0.09 & $-$0.02 & +0.11 & $-$0.13 & $-$0.07  & Paper~I \\
DX Del(2)&  2453184.980 &  6340 &  2.63 &  3.0 & $-$0.26 & $-$0.04 & $-$0.06 & +0.06 & $-$0.07 & +0.03  & Paper~I \\
DX Del(3)&  2453185.069 &  6260 &  2.69 &  3.4 & $-$0.22 & $-$0.01 & $-$0.18 & +0.12 & $-$0.06 & +0.17  & Paper~I \\
IO Lyr   &  2453184.902 &  6350 &  2.11 &  2.8 & $-$1.39 & $-$0.47 &($-$0.38)& +0.54 & $-$0.37 &($-$0.09) & \\
KX Lyr   &  2453184.910 &  6580 &  2.91 &  2.8 & $-$0.30 & $-$0.09 & +0.21 & +0.11 & +0.00 & $-$0.30  & \\
RR Cet   &  2453185.083 &  6720 &  2.36 &  2.7 & $-$1.33 & $-$0.27 & +0.48 & +0.72 & $-$0.12 & $-$0.76  & \\
RR Lyr(1)&  2453184.796 &  7530 &  2.80 &  1.9 & $-$1.31 & $-$1.03 & +0.27 & +0.81 & $-$0.03 & $-$1.30  & Paper~I \\
RR Lyr(2)&  2453184.882 &  6710 &  2.36 &  3.1 & $-$1.41 & $-$0.47 & +0.21 & +0.62 & $-$0.17 & $-$0.68  & Paper~I \\
RR Lyr(3)&  2453184.885 &  6620 &  2.16 &  3.0 & $-$1.45 & $-$0.53 & +0.33 & +0.66 & $-$0.24 & $-$0.86  & Paper~I \\
RR Lyr(4)&  2453184.971 &  6210 &  2.14 &  2.8 & $-$1.45 & $-$0.46 & +0.47 & +0.62 & $-$0.29 & $-$0.93  & Paper~I \\
RR Lyr(5)&  2453185.056 &  6040 &  2.09 &  3.0 & $-$1.50 & $-$0.64 &($-$0.13)& +0.62 & $-$0.19 & ($-$0.51)  & Paper~I \\
RS Boo   &  2453184.740 &  6630 &  2.76 &  2.7 & $-$0.23 & $-$0.01 & +0.18 & +0.14 & +0.00 & $-$0.19  & \\
SW And   &  2453185.018 &  6750 &  2.60 &  3.3 & $-$0.21 & $-$0.05 & $-$0.12 & +0.13 & $-$0.07 & +0.08  & \\
TV Lib   &  2453184.935 &  6580 &  2.80 &  2.2 & $-$0.48 & $-$0.12 & +0.05 & +0.36 & +0.12 & $-$0.17  & \\
TW Her   &  2453184.814 &  7550 &  2.59 &  4.4 & $-$0.48 & $-$0.10 & +0.37 & +0.30 & +0.06 & $-$0.47  & \\
UU Cet   &  2453185.077 &  6180 &  2.58 &  3.5 & $-$1.28 & $\cdots$ & +0.65 & +0.48 & $-$0.47 & $\cdots$  & \\
V413 Oph &  2453184.864 &  7040 &  2.55 &  3.2 & $-$0.87 & +0.05 & +0.28 & +0.68 & +0.06 & $-$0.24  & \\
V440 Sgr &  2453184.878 &  7660 &  2.73 &  3.5 & $-$1.16 & $-$0.18 & +0.03 & +0.62 & $-$0.18 & $-$0.22  & \\
VX Her   &  2453184.805 &  6250 &  2.33 &  3.0 & $-$1.46 & $-$0.09 & $\cdots$ & +0.59 & $-$0.08 & $\cdots$  & \\
VY Ser(1)&  2453184.761 &  5970 &  1.96 &  2.4 & $-$1.86 & $\cdots$ & +0.53 & +0.73 & $-$0.37 & $\cdots$  & Paper~I \\
VY Ser(2)&  2453184.845 &  5990 &  2.17 &  2.7 & $-$1.82 &($-$0.66)& +0.83 & +0.78 & $-$0.23 & ($-$1.49)  & Paper~I \\
VY Ser(3)&  2453184.943 &  6170 &  2.50 &  3.0 & $-$1.67 &($-$0.55)&(+0.08)& +0.72 & $-$0.16 & ($-$0.63)  & Paper~I \\
VY Ser(4)&  2453185.002 &  6210 &  2.44 &  3.4 & $-$1.63 &($-$0.46)& +0.98 & +0.65 & $-$0.64 & ($-$1.43)  & Paper~I \\
Procyon  &        $\cdots$ &  6612 &  4.00 &  2.0 & $-$0.02 & +0.07 & +0.12 & +0.08 & +0.10 & $-$0.05  & \\
Sun      &        $\cdots$ &  5780 &  4.44 &  1.0 &  0.00 &  0.00 &  0.00 &  0.00 &  0.00 &  0.00  & \\
\hline
$\langle$AO Peg$\rangle$ & $\cdots$ & $\cdots$ & $\cdots$ & $\cdots$& $-$1.19 & $-$0.47 &($-$0.09)& +0.70 & $-$0.54 &($-$0.38) & \\
$\langle$DH Peg$\rangle$ & $\cdots$ & $\cdots$ & $\cdots$ & $\cdots$& $-$1.16 & $-$0.27 & +0.76 & +0.54 & +0.03 & $-$1.03  & \\
$\langle$DX Del$\rangle$ & $\cdots$ & $\cdots$ & $\cdots$ & $\cdots$& $-$0.23 & $-$0.05 & $-$0.06 & +0.09 & $-$0.08 & +0.01  & \\
$\langle$RR Lyr$\rangle$ & $\cdots$ & $\cdots$ & $\cdots$ & $\cdots$& $-$1.42 & $-$0.54 & +0.29 & +0.65 & $-$0.20 & $-$0.83  & \\
$\langle$VY Ser$\rangle$ & $\cdots$ & $\cdots$ & $\cdots$ & $\cdots$& $-$1.75 & ($-$0.57) &(+0.78)& +0.74 & $-$0.29 &($-$1.35) & \\
\hline
\end{tabular}
\end{center}
(1) Object name. In case that several spectra are available for a star, an integer (enclosed by parenthesis) 
is attached to indicate the sequence number. (2) Heliocentric Julian date (in day) at the time of observation.
(3) Effective temperature (in K). (4) Logarithmic surface gravity (in dex; $g$ is in cm~s$^{-2}$).
(5) Microturbulent velocity dispersion (in km~s$^{-1}$). (6) Logarithmic Fe abundance relative to the Sun (in dex).
(7) Logarithmic C-to-Fe abundance ratio (in dex). (8) Logarithmic N-to-Fe abundance ratio (in dex).
(9) Logarithmic O-to-Fe abundance ratio (in dex). (10) Logarithmic Na-to-Fe abundance ratio (in dex).
(11) Logarithmic C-to-N abundance ratio (in dex). (12) Additional remark.\\
See footnote~1 in Sect.~1 for the definition of [X/H] or [X/Y] (X or Y denote any element). 
Non-LTE abundances were used for all the quantities related to C, N, O, and Na, while [Fe/H]
($\equiv A({\rm Fe}) - 7.60$) was derived with LTE, where $A$(Fe) is the Fe abundance of a star 
determined in Sect.~3 and 7.60 is the adopted solar Fe abundance (see Sect.~3.1 in Paper~I). 
Note that the parenthesised values are unreliable and 
should not be seriously taken (cf. the last paragraph of Sect.~4.3). The averaged data for the five stars 
(for each of which several spectra are available) are summarised in the last five rows. 
The atmospheric parameters of 17 objects/spectra (indicated as ``Paper~I'' in column~12) were derived 
already in Paper~I and thus used unchanged.

\end{minipage}
\end{table*}
\section{Atmospheric parameters}

The atmospheric parameters [$T_{\rm eff}$ (effective temperature), $\log g$ (surface 
gravity), $v_{\rm t}$ (microturbulence), and [Fe/H] (metallicity)] for each RR~Lyr star, 
which are necessary to construct an adequate model atmosphere used for abundance 
determination, have to be established from the spectrum (itself) to be analysed,
because the physical condition of a pulsating star is time-variable.

In Paper~I, these parameters were determined for the 15 spectra of 4 stars based on the 
conventional method using Fe~{\sc i} and Fe~{\sc ii} lines, which requires that three 
conditions be simultaneously fulfilled: (i) excitation equilibrium of Fe~{\sc i} 
(abundances do not depend upon the excitation potential), (ii) ionisation equilibrium 
between Fe~{\sc i} and Fe~{\sc ii} (equality of the mean abundances), and curve-of-growth 
matching (abundances do not depend upon the line strengths).
Following the same procedure as adopted in Paper~I (cf. Sect.~3 therein),
these parameters were newly determined for the 19 spectra of 18 stars (out of 34 spectra
for 22 stars in total) which were not touched in Paper~I.
The final parameter solutions are given in Table~1, and the resulting Fe abundances 
for Fe~{\sc i} and Fe~{\sc ii} lines (the detailed line-by-line data are presented
in ``felines.dat'' of the online material) are plotted against $W_{\lambda}$ (equivalent 
width) and $\chi_{\rm low}$ (lower excitation potential) in Fig.~1. 

\setcounter{figure}{0}
\begin{figure}
\begin{minipage}{80mm}
\includegraphics[width=8.0cm]{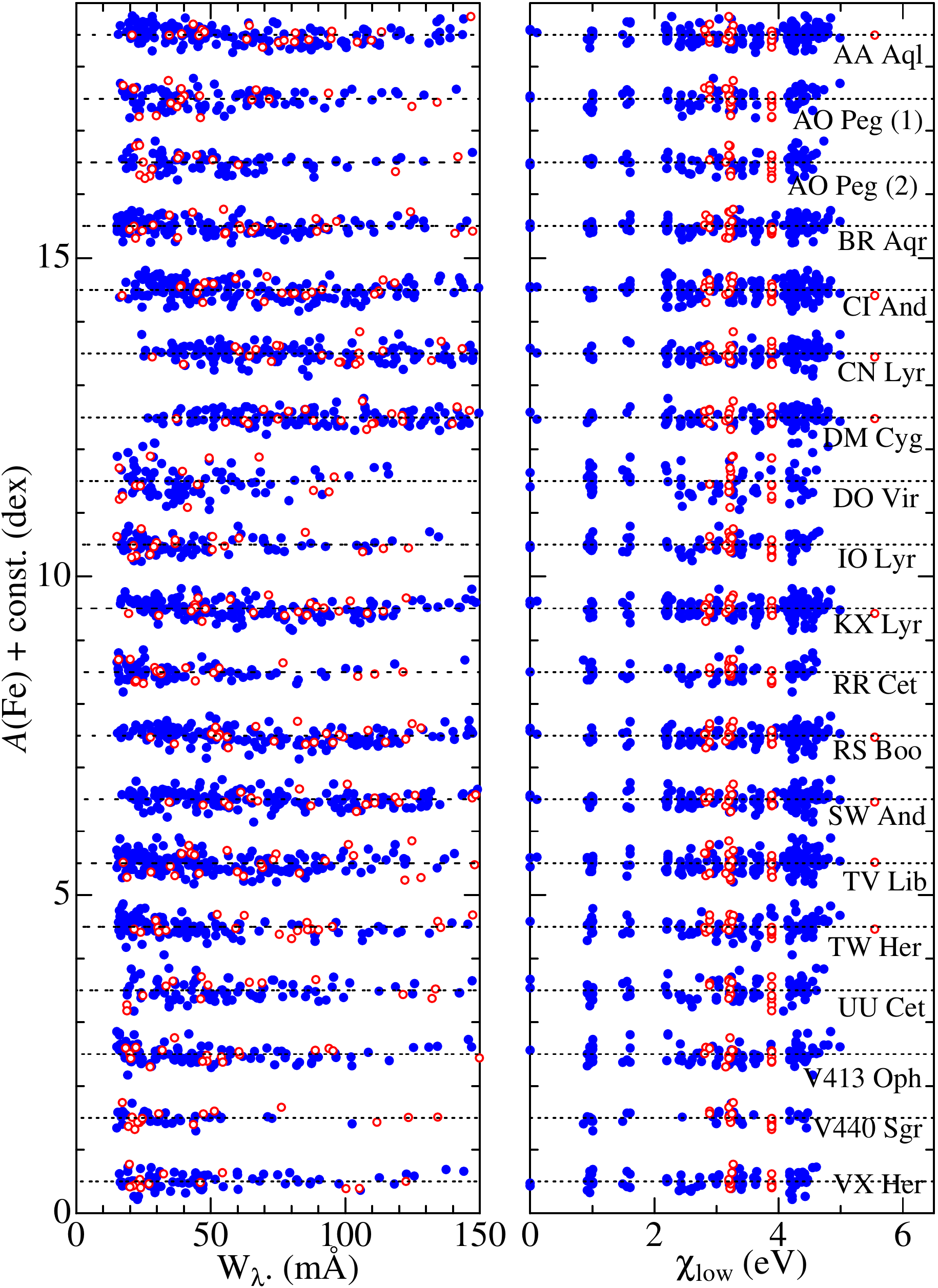}
\caption{
Fe abundance vs. equivalent width relation (left panel) 
and Fe abundance vs. lower excitation potential relation
(right panel) corresponding to the finally established
atmospheric parameters of $T_{\rm eff}$, $\log g$, and $v_{\rm t}$
for each of the 19 spectra of 18 RR~Lyr stars (which were not used 
in Paper~I and newly analysed in this study). The filled and open circles 
correspond to Fe~{\sc i} and Fe~{\sc ii} lines, respectively. 
The results for each stars are shown relative to the mean abundances 
indicated by the horizontal dashed lines, and vertically offset by 1.0 
relative to the adjacent ones. (This figure is arranged in the same manner
as in Fig.~3 of Paper~I, where the results for the other 15 spectra of 
4 stars are depicted.)
}
\label{fig1}
\end{minipage}
\end{figure}

In this connection, the parameters adopted in Paper~II were determined (for 36 spectra
of 23 stars) based on the same principle [requirements (i), (ii), and (iii)] but using 
a different computer program (independently developed by the Chinese group) along with 
different set of Fe lines. These Paper~II values of $T_{\rm eff}$, $\log g$, 
$v_{\rm t}$, and [Fe/H] are compared with those of Paper~I and those newly established for this study 
in Fig.~2a--d, where we can see that both are more or less consistent with each other
except for a few cases (e.g., $\log g$ of DO~Vir or $v_{\rm t}$ of TW~Her).
An important difference is, however, that parameter determination for V445~Oph 
(somehow accomplished in Paper~II) unfortunately failed in this study, because convergence 
could not be attained with anomalously large Fe abundances. Accordingly, V445~Oph (two 
spectra) had to be excluded from our analysis, which is the reason for the slight difference
in the programme stars between this study (cf. Table~1) and Paper~II (cf. Table~1 therein).

The model atmosphere for each star (spectrum) was then constructed by three-dimensionally 
interpolating Kurucz's (1993) ATLAS9 model grid (for $v_{\rm t} = 2$~km~s$^{-1}$) 
in terms of $T_{\rm eff}$, $\log g$, and [Fe/H].
The non-LTE departure coefficients for C, N, O, and Na were also computed for a grid of
144 models resulting from combinations of six $T_{\rm eff}$ values (5500, 6000, 6500, 7000,
7500, and 8000~K), four $\log g$ values (1.5, 2.0, 2,5, and 3.0), and six [Fe/H] values 
($-2.0$, $-1.5$, $-1.0$, $-0.5$, 0.0, and  +0.5),
which were further interpolated for each star as done for model atmospheres. 
Regarding more details about non-LTE calculations, see Takeda \& Honda (2005) (for C, N, and O) 
and Takeda et al.(2003) (for Na) and the references therein.

\setcounter{figure}{1}
\begin{figure}
\begin{minipage}{80mm}
\includegraphics[width=8.0cm]{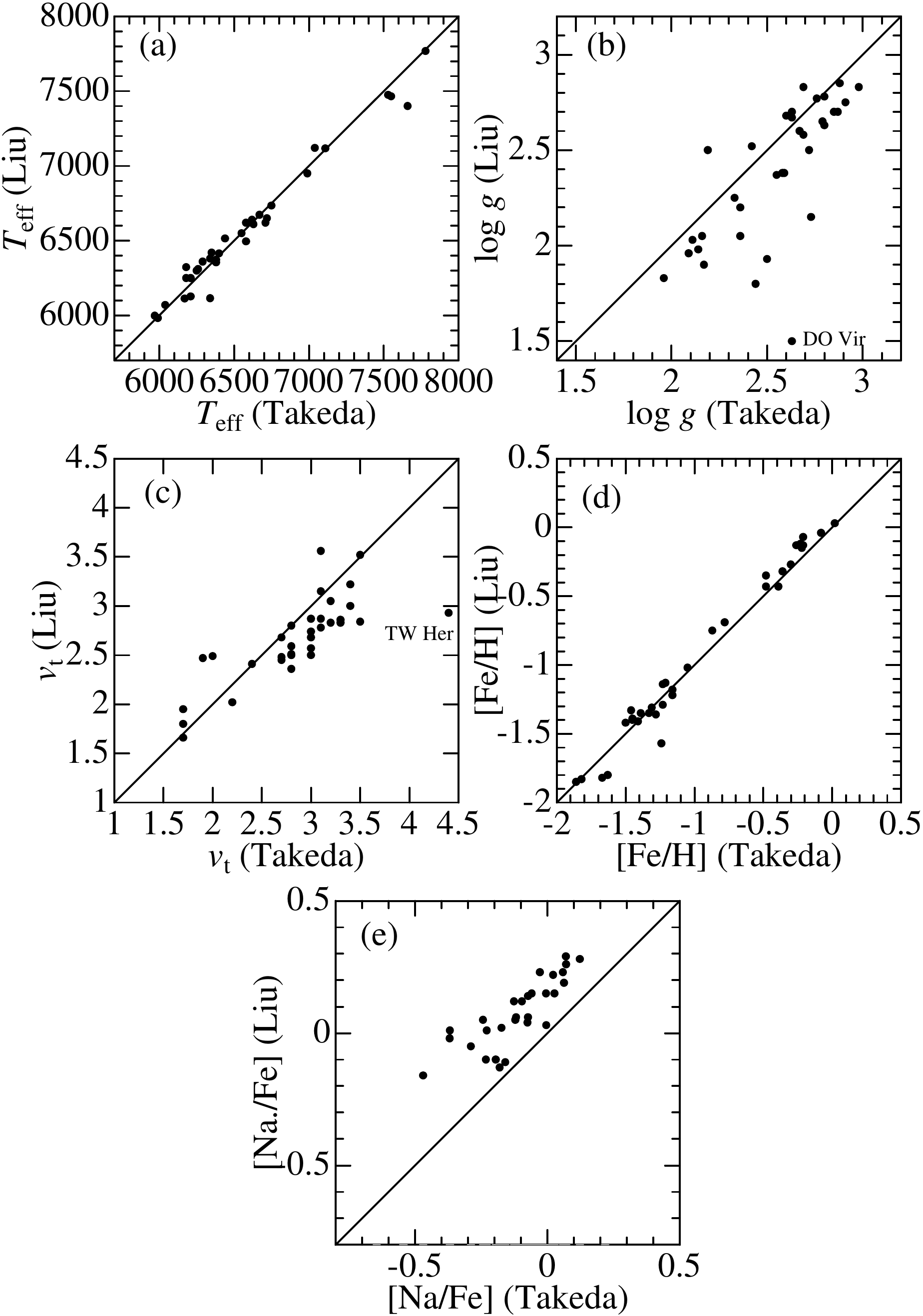}
\caption{Comparison of the atmospheric parameters (and Na abundances) 
determined in Paper~I and this study (abscissa) with 
those independently derived in Paper~II (Liu et al. 2013) based on the same 
spectra (ordinate). (a) $T_{\rm eff}$, (b) $\log g$, (c) $v_{\rm t}$,
(d) [Fe/H], and (e) [Na/Fe] (note that Liu et al.'s Na abundances were derived 
based on the assumption of LTE, while non-LTE corrections are taken into 
consideration in this study). 
}
\label{fig2}
\end{minipage}
\end{figure}

\section{Abundance determination} 

In the differential analysis (relative to the Sun) adopted in this study, 
only one specific line feature of high quality (e.g., almost free from blending of 
other lines, unaffected by any telluric lines) is used for determining the abundance 
of each element: C~{\sc i} 5380 line (for C), N~{\sc i} 7468 line (for N), O~{\sc i} 
7771/7774/7775 triplet lines (for O), and Na~{\sc i} 5682/5688 doublet lines (for Na).
The determination procedures of abundances and related quantities (e.g., 
non-LTE correction, uncertainties due to ambiguities of atmospheric parameters) 
consist of two consecutive steps.

\subsection{Synthetic spectrum fitting} 

The first step is to find the solutions for the non-LTE abundances of relevant elements 
($A_{1}^{\rm N}, A_{2}^{\rm N}, \ldots$), macrobroadening velocity ($v_{\rm M}$),\footnote{
This $v_{\rm M}$ is the $e$-folding half width of the Gaussian macrobroadening function 
to be convolved, $f_{\rm M}(v) \propto \exp{[-(v/v_{\rm M})^{2}]}$,
which represents the combined effects of rotational broadening, macroturbulence,
and instrumental broadening.}
and radial velocity ($V_{\rm rad}$) such as those accomplishing the
best fit (minimising $O-C$ residuals) between theoretical and 
observed spectra, while applying the automatic fitting algorithm 
(Takeda 1995). Four wavelength regions were selected for this purpose:
(1) 5378--5382~\AA\ region (for C), (2) 7460--7470~\AA\ region (for N),  
(3) 7770--7777~\AA\ region (for O), and 5681--5689.5~\AA\ region (for Na). 
More information about this fitting analysis (varied elemental abundances, used 
data of atomic lines, etc.) is summarised in Table~2, and the atomic data of important
lines are presented in Table~3.
How the theoretical spectrum for the converged solutions fits well 
with the observed spectrum is displayed in Figs.~3--6 for each region. 

\setcounter{figure}{2}
\begin{figure}
\begin{minipage}{80mm}
\includegraphics[width=8.0cm]{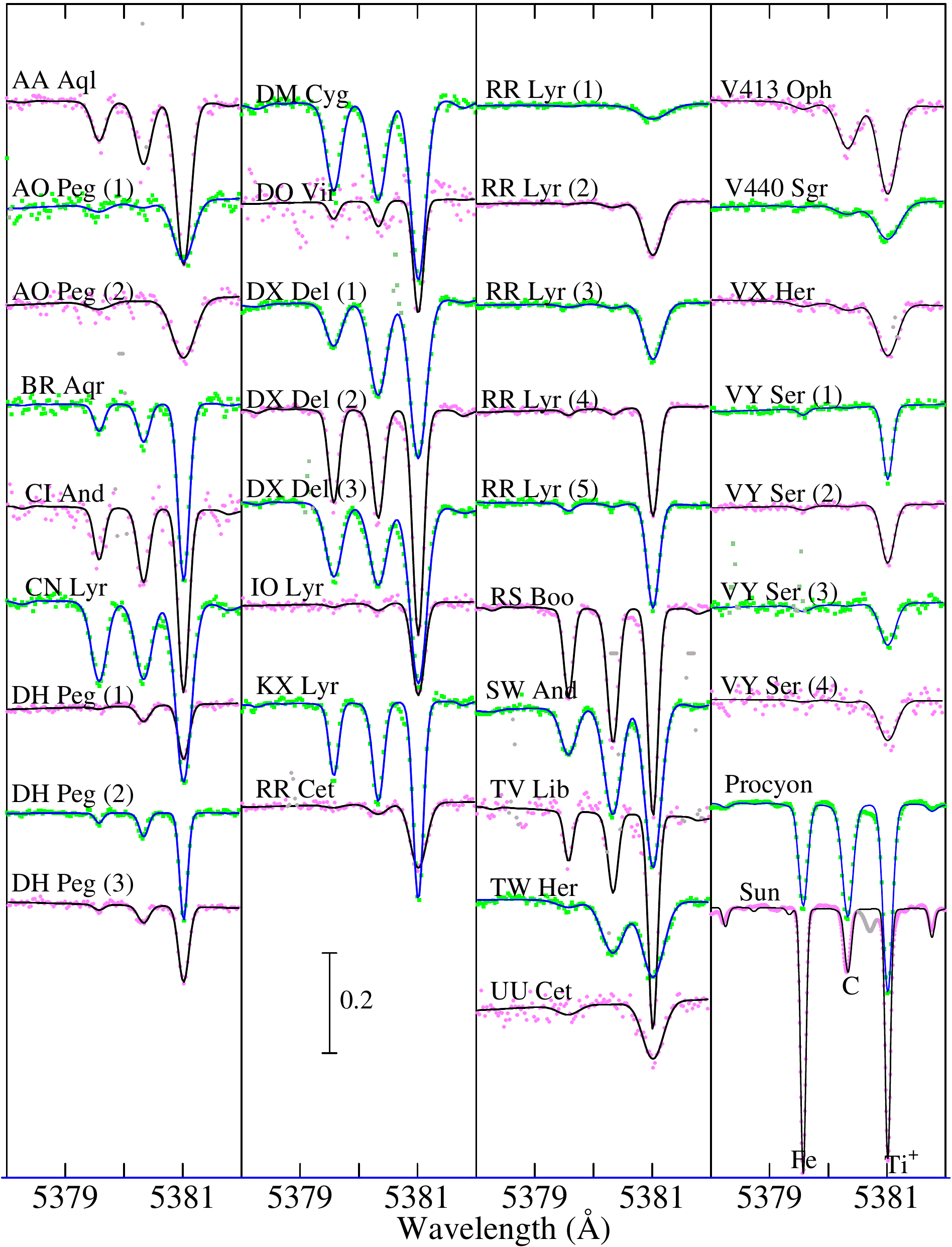}
\caption{
Synthetic spectrum fitting in the 5378--5382~\AA\ region 
comprising the C~{\sc i}~5380 line. 
The best-fit theoretical spectra are shown by solid lines. 
The observed data are plotted by symbols, where those used 
in the fitting are coloured in pink or green, while those rejected 
in the fitting (e.g., due to spectrum defect) are depicted in gray.
In each panel, the spectra are arranged in the same order 
as in Table~1, and a vertical offset of 0.2 
is applied to each spectrum relative to the adjacent one. 
}
\label{fig3}
\end{minipage}
\end{figure}

\setcounter{figure}{3}
\begin{figure}
\begin{minipage}{80mm}
\includegraphics[width=8.0cm]{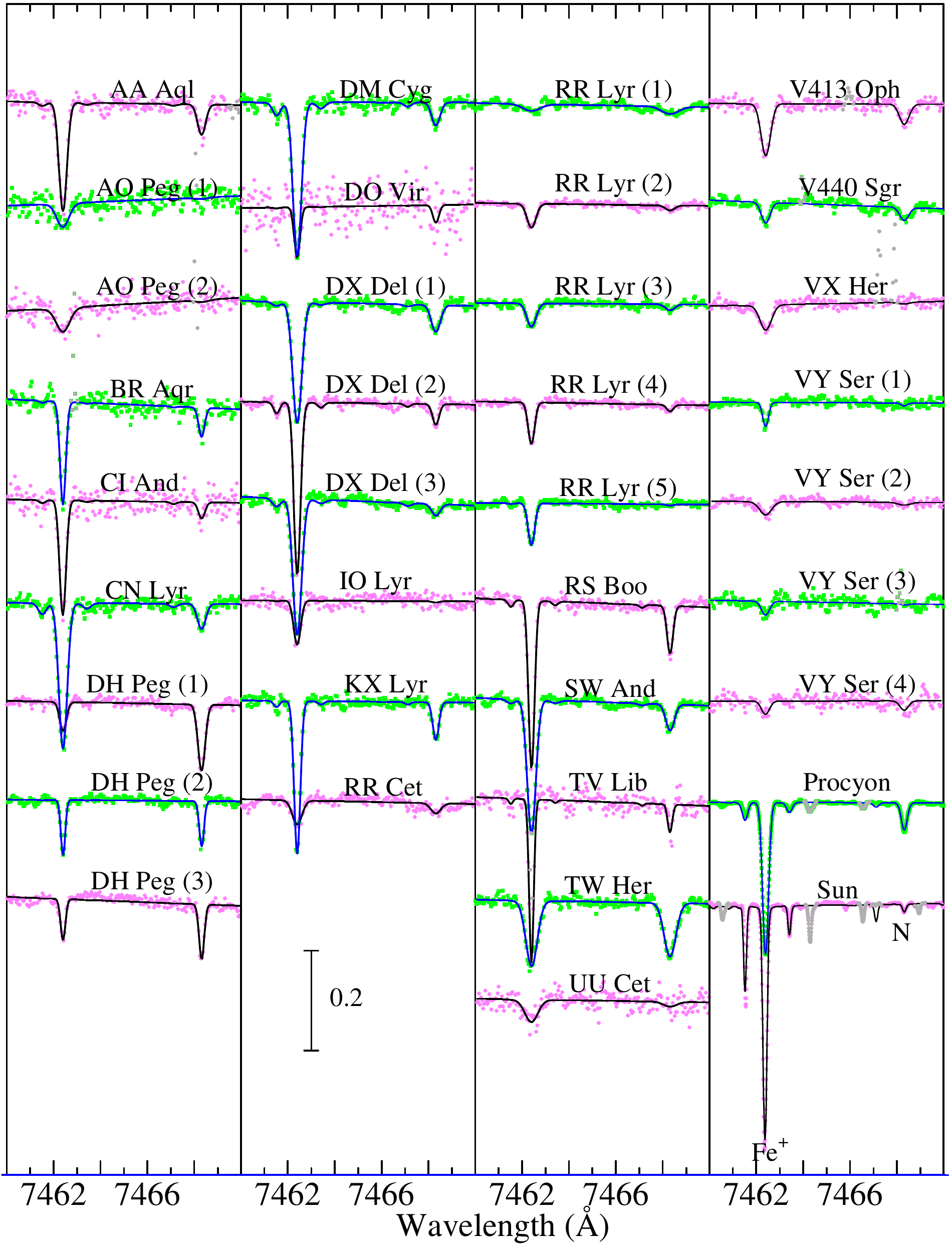}
\caption{
Synthetic spectrum fitting in the 7460--7470~\AA\ region 
comprising the N~{\sc i}~7468 line. Otherwise, the same as in Fig.~3. 
}
\label{fig4}
\end{minipage}
\end{figure}

\setcounter{figure}{4}
\begin{figure}
\begin{minipage}{80mm}
\includegraphics[width=8.0cm]{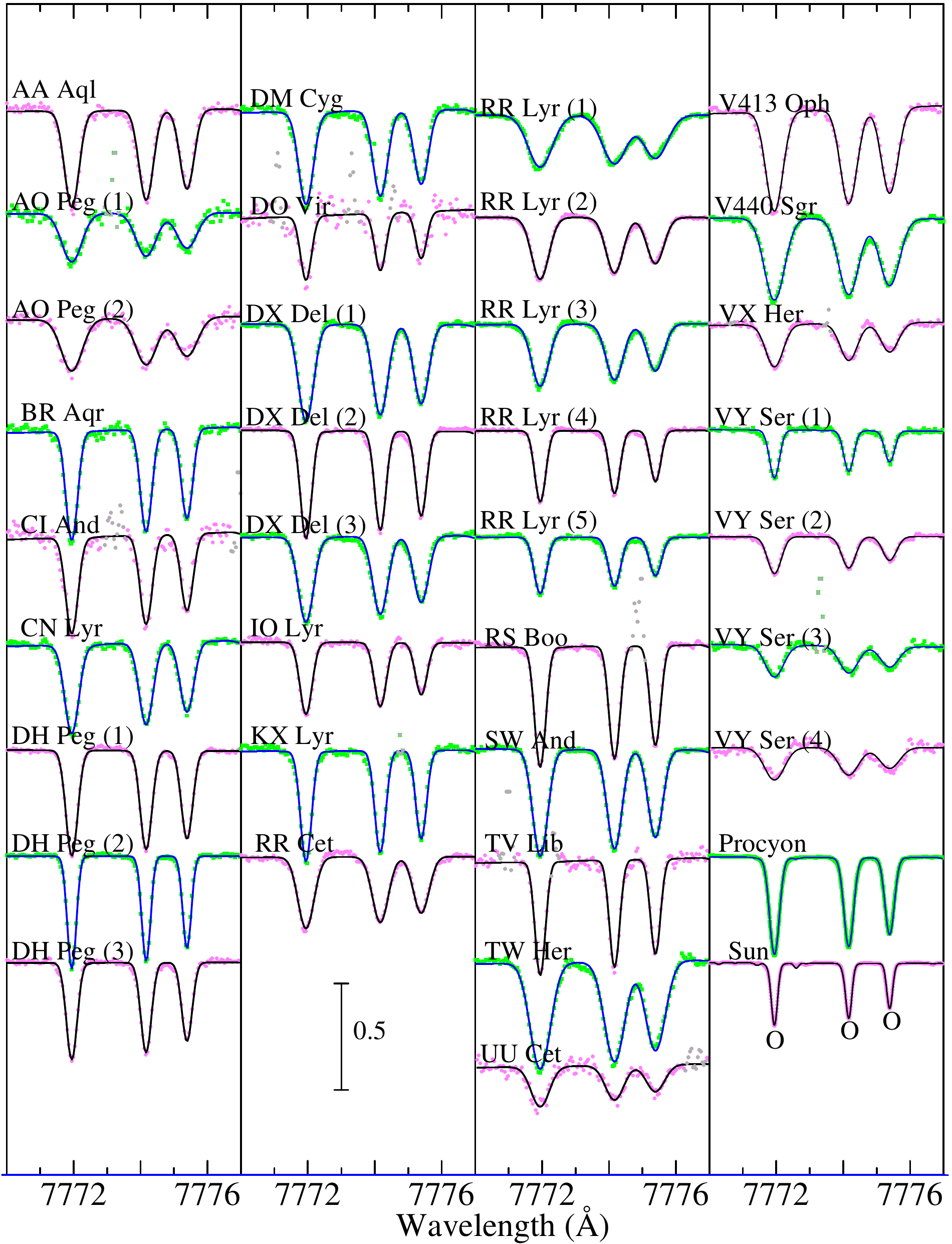}
\caption{
Synthetic spectrum fitting in the 7770--7777~\AA\ region 
comprising the O~{\sc i}~7771/7774/7775 lines.
A vertical offset of 0.5 is applied to each spectrum relative 
to the adjacent one. Otherwise, the same as in Fig.~3. 
}
\label{fig5}
\end{minipage}
\end{figure}

\setcounter{figure}{5}
\begin{figure}
\begin{minipage}{80mm}
\includegraphics[width=8.0cm]{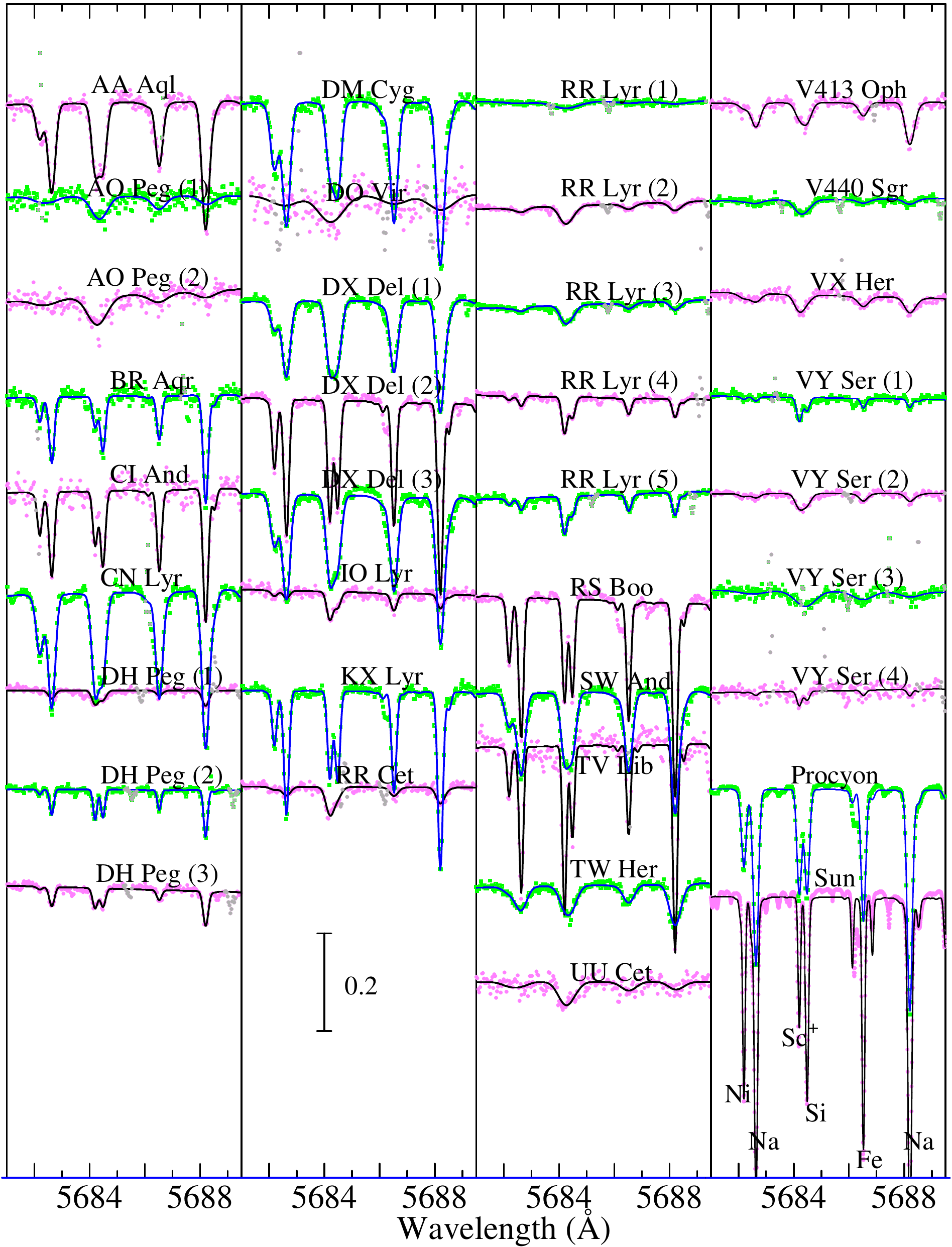}
\caption{
Synthetic spectrum fitting in the 5681--5689.5~\AA\ region 
comprising the Na~{\sc i}~5682/5688 lines. Otherwise, 
the same as in Fig.~3. 
}
\label{fig6}
\end{minipage}
\end{figure}

\setcounter{table}{1}
\begin{table*}
\begin{minipage}{180mm}
\small
\caption{Outline of spectrum-fitting analysis in this study.}
\begin{center}
\begin{tabular}{ccccc}\hline\hline
Purpose & Fitting range (\AA) & Abundances varied$^{*}$ & Atomic data source & Figure \\
\hline
C abundance from C~{\sc i} 5380 & 5378--5382 & C, Ti, Fe & KB95m1 & Fig.~3 \\
N abundance from N~{\sc i} 7468  & 7460--7470 & N, Fe & KB95m2 & Fig.~4 \\
O abundance from O~{\sc i} 7771--5  & 7770--7777 & O & KB95m3 & Fig.~5 \\
Na abundance from Na~{\sc i} 5682/5688 & 5681--5689.5 & Na, Si, Sc, Fe, Ni & KB95 & Fig.~6 \\ 
\hline
\end{tabular}
\end{center}
$^{*}$ The abundances of other elements than these were fixed in the fitting. \\
KB95m1 --- All the atomic line data presented in Kurucz \& Bell (1995) were used, 
excepting that Ti~{\sc ii} 5379.15 and Fe~{\sc i} 5382.47 were neglected (by 
drastically suppressing their $gf$ values to a negligible level). \\
KB95m2 --- All the atomic line data were taken from Kurucz \& Bell (1995), excepting
that S~{\sc i} 7468.59 was neglected. \\
KB95m3 --- All the atomic line data presented in Kurucz \& Bell (1995) were used, 
excepting that the $\log gf$ values of four Fe~{\sc i} lines (at 7770.28, 7771.43, 7772.60, 
and 7774.00~\AA) were empirically adjusted (as $-5.12$, $-2.48$, $-2.16$, and $-2.27$,
respectively) while four lines (Ca~{\sc i} 7771.24, 7775.50, 7775.76, and Ti~{\sc i} 7773.90) 
were neglected (cf. Sect.~3.3 in Takeda, Kawanomoto \& Sadakane 1998).\\
KB95 --- All the atomic line data given in Kurucz \& Bell (1995) were used unchanged.
\end{minipage}
\end{table*}

\setcounter{table}{2}
\begin{table*}
\begin{minipage}{180mm}
\small
\caption{Adopted atomic data of the relevant C, N, O, and Na lines.}
\begin{center}
\begin{tabular}{cccccccc}\hline\hline
Line & Multiplet & $\lambda$ & $\chi_{\rm low}$ & $\log gf$ & Gammar & Gammas & Gammaw\\
     &  No.      & (\AA) & (eV) & (dex) & (dex) & (dex) & (dex) \\  
\hline
C~{\sc i} 5380 & (11) &  5380.337 & 7.685 & $-1.842$ & (7.89) & $-$4.66 & ($-$7.36)\\
\hline
N~{\sc i} 7468 & (3)  & 7468.312 & 10.336 & $-0.270$  & 8.64 & $-$5.40 & ($-$7.60)\\
\hline
O~{\sc i} 7771 & (1) & 7771.944 & 9.146 & $+0.324$ & 7.52 & $-$5.55 & ($-$7.65)\\
O~{\sc i} 7774 & (1) & 7774.166 & 9.146 & $+0.174$ & 7.52 & $-$5.55 & ($-$7.65)\\
O~{\sc i} 7775 & (1) & 7775.388 & 9.146 & $-0.046$ & 7.52 & $-$5.55 & ($-$7.65)\\
\hline
Na~{\sc i} 5682 & (6) & 5682.633 & 2.102 & $-0.700$ & (7.84) & ($-$5.00) & ($-$7.23)\\
Na~{\sc i} 5688 & (6) & 5688.204 & 2.104 & $-0.404$ & (7.84) & ($-$5.00) & ($-$7.23)\\
\hline
\end{tabular}
\end{center}
Following the multiplet number, laboratory (air) wavelength, lower excitation potential, and 
$gf$ value in columns 2--5, three kinds of damping parameters are presented in columns 6--8:  
Gammar is the radiation damping width (s$^{-1}$) [$\log\gamma_{\rm rad}$], 
Gammas is the Stark damping width (s$^{-1}$) per electron density (cm$^{-3}$) 
at $10^{4}$ K [$\log(\gamma_{\rm e}/N_{\rm e})$], and
Gammaw is the van der Waals damping width (s$^{-1}$) per hydrogen density 
(cm$^{-3}$) at $10^{4}$ K [$\log(\gamma_{\rm w}/N_{\rm H})$]. \\
All the data were taken from Kurucz \& Bell (1995), except for 
the parenthesised damping parameters (unavailable in their compilation), 
for which the default values computed by the WIDTH9 program were assigned.
\end{minipage}
\end{table*}

\subsection{Abundances from equivalent widths} 

As the second step, with the help of Kurucz's (1993) WIDTH9 program 
(which had been considerably modified in various respects; e.g., 
inclusion of non-LTE effects, treatment of total equivalent width for 
multi-component lines; etc.), the equivalent widths ($W$) of the representative 
lines were ``inversely'' computed from the abundance solutions (resulting from 
spectrum synthesis) along with the adopted atmospheric model/parameters; i.e., 
$W_{5380}$ (for C~{\sc i} 5380), $W_{7468}$ (for N~{\sc i} 7468), $W_{7771}$ 
(for O~{\sc i} 7771), and $W_{5688}$ (for Na~{\sc i} 5688), because they are 
easier to handle in practice (e.g., for estimating uncertainties due to errors 
in atmospheric parameters).
Such derived $W$ values were then analysed by using WIDTH9 to determine 
$A^{\rm N}$ (NLTE abundance) and $A^{\rm L}$ (LTE abundance), from 
which the NLTE correction $\Delta (\equiv A^{\rm N} - A^{\rm L})$
was further derived. Since the Sun is adopted as the standard star of
abundance reference, the relative abundance is defined as [X/H] $\equiv$ 
$A_{*}^{\rm N}$(X) $-$ $A_{\odot}^{\rm N}$(X), from which the X-to-Fe 
ratio is derived as [X/Fe] $\equiv$ [X/H] $-$ [Fe/H] (X = C, N, O, and Na).
The resulting [C/Fe], [N/Fe], [O/Fe], and [Na/Fe] for each star are given in Table~1 
(more complete results including $W$ and $\Delta$ are separately presented in 
``abunds.dat'' of the supplementary material). 
Figs. 7(C), 8(N), 9(O), and 10(Na) graphically show the equivalent width ($W$),
non-LTE correction ($\Delta$), non-LTE abundance ($A^{\rm N}$), 
and abundance variations in response to parameter changes (see the 
following Sect.~4.3), as functions of $T_{\rm eff}$.

\setcounter{figure}{6}
\begin{figure}
\begin{minipage}{80mm}
\includegraphics[width=8.0cm]{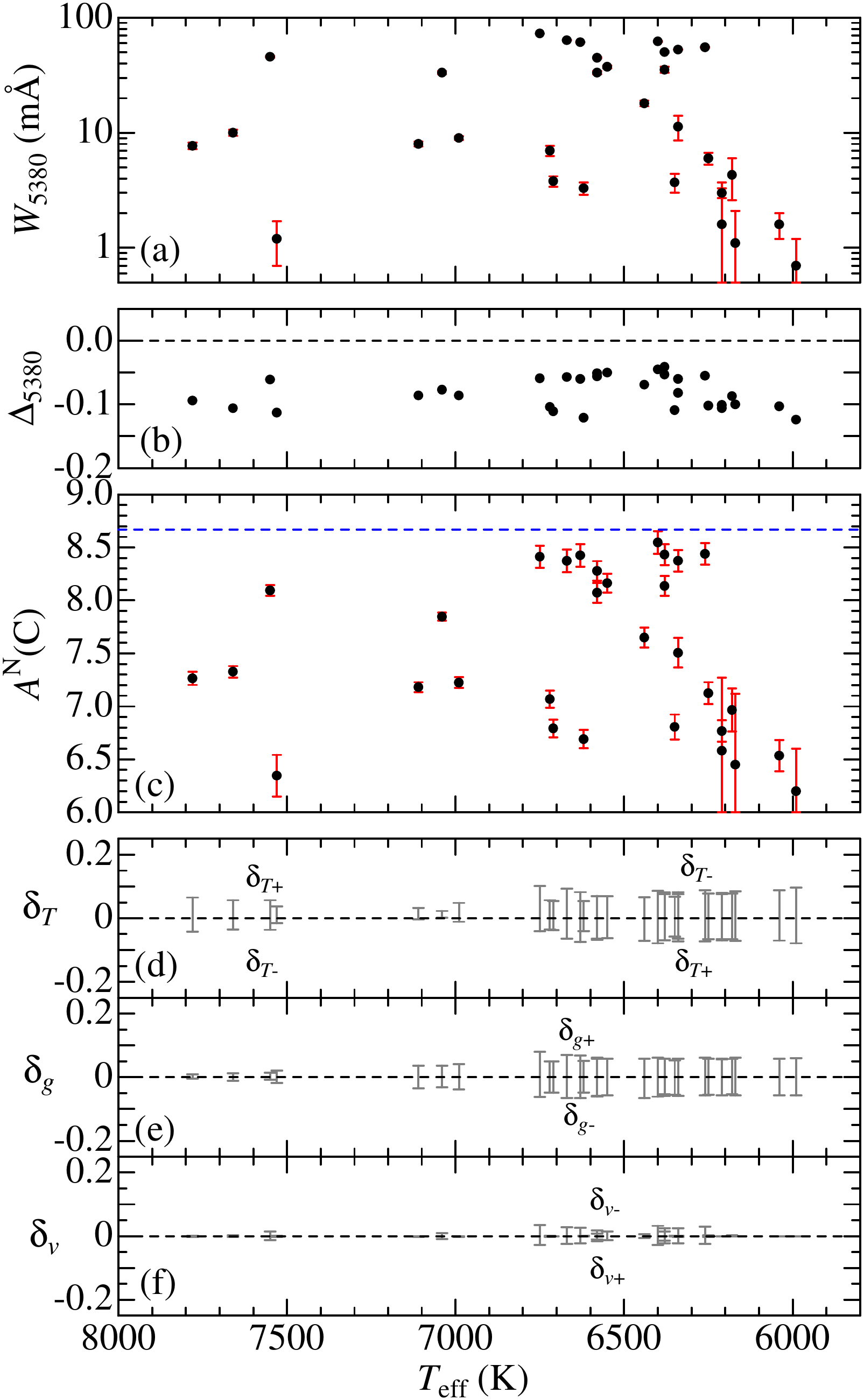}
\caption{
Carbon abundance and C~{\sc i}~5380-related quantities 
plotted against $T_{\rm eff}$. 
(a) $W_{5380}$ (equivalent width of C~{\sc i} 5380, where the error bars
attached to each symbol denote $\pm \delta W$; cf. Sect.~4.3), 
(b) $\Delta_{5380}$ (non-LTE correction for C~{\sc i} 5380),
(c) $A^{\rm N}$(C) (non-LTE abundance derived from C~{\sc i} 5380)
where the error bar denotes $\pm\delta W$ defined in Sect.~4.3,
(d) $\delta_{T+}$ and $\delta_{T-}$ (abundance variations 
in response to $T_{\rm eff}$ changes of +200~K and $-200$~K), 
(e) $\delta_{g+}$ and $\delta_{g-}$ (abundance variations 
in response to $\log g$ changes by $+0.2$~dex and $-0.2$~dex), 
and (f) $\delta_{v+}$ and $\delta_{v-}$ (abundance 
variations in response to perturbing the $v_{\rm t}$ value
by +20\% and $-$20\%).
The solar $A^{\rm N}_{\odot}$(C) value of 8.67, which is adopted as the reference, 
is indicated by the horizontal dashed line in panel (c). 
Note that the ordinate in panel (a) is expressed in the logarithmic scale.
}
\label{fig7}
\end{minipage}
\end{figure}

\setcounter{figure}{7}
\begin{figure}
\begin{minipage}{80mm}
\includegraphics[width=8.0cm]{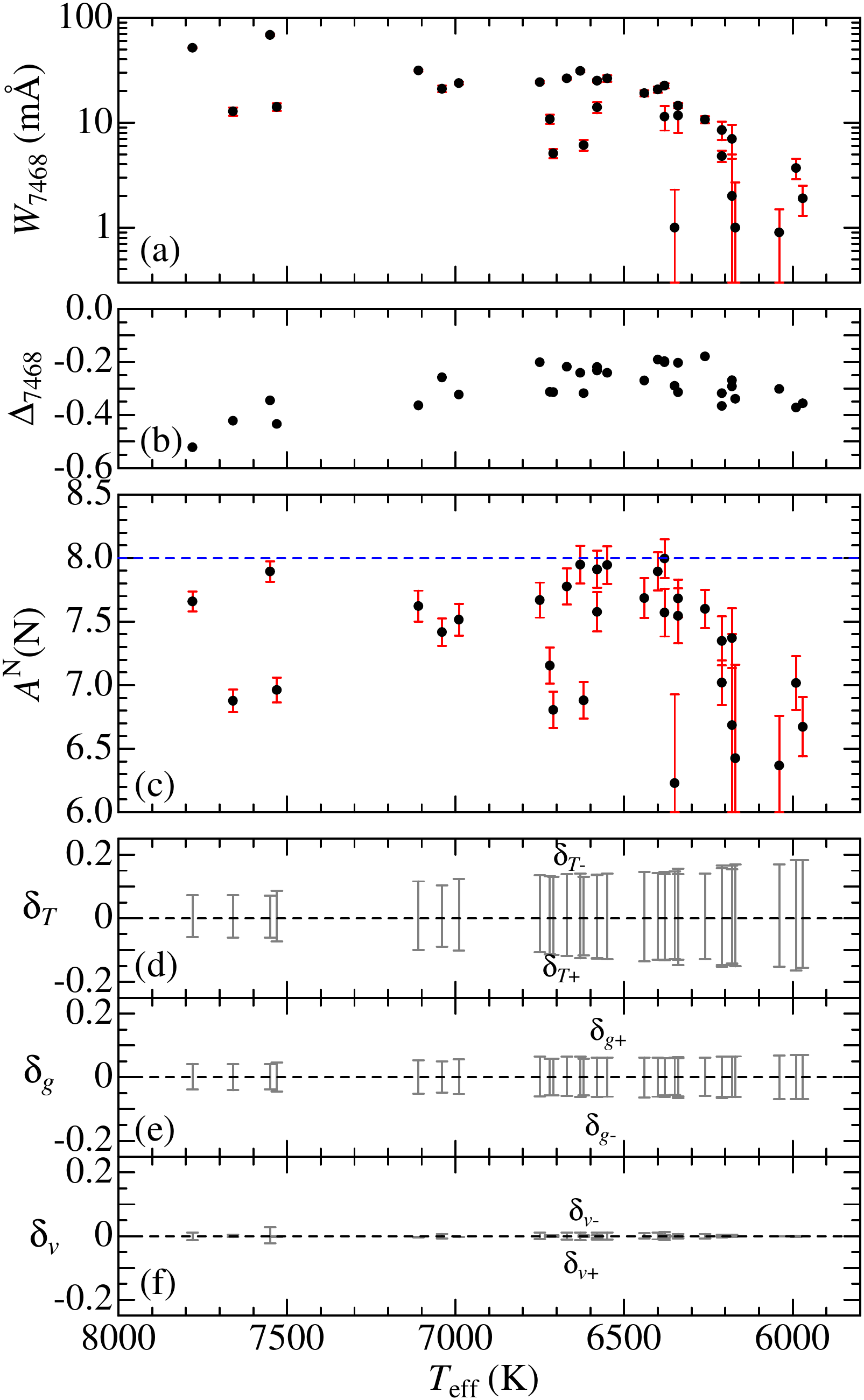}
\caption{
Nitrogen abundance and N~{\sc i}~7468-related quantities 
plotted against $T_{\rm eff}$. 
The reference solar $A^{\rm N}_{\odot}$(N) value of 8.00 is indicated 
by the horizontal dashed line in panel (c).
Otherwise, the same as in Fig.~7. 
}
\label{fig8}
\end{minipage}
\end{figure}

\setcounter{figure}{8}
\begin{figure}
\begin{minipage}{80mm}
\includegraphics[width=8.0cm]{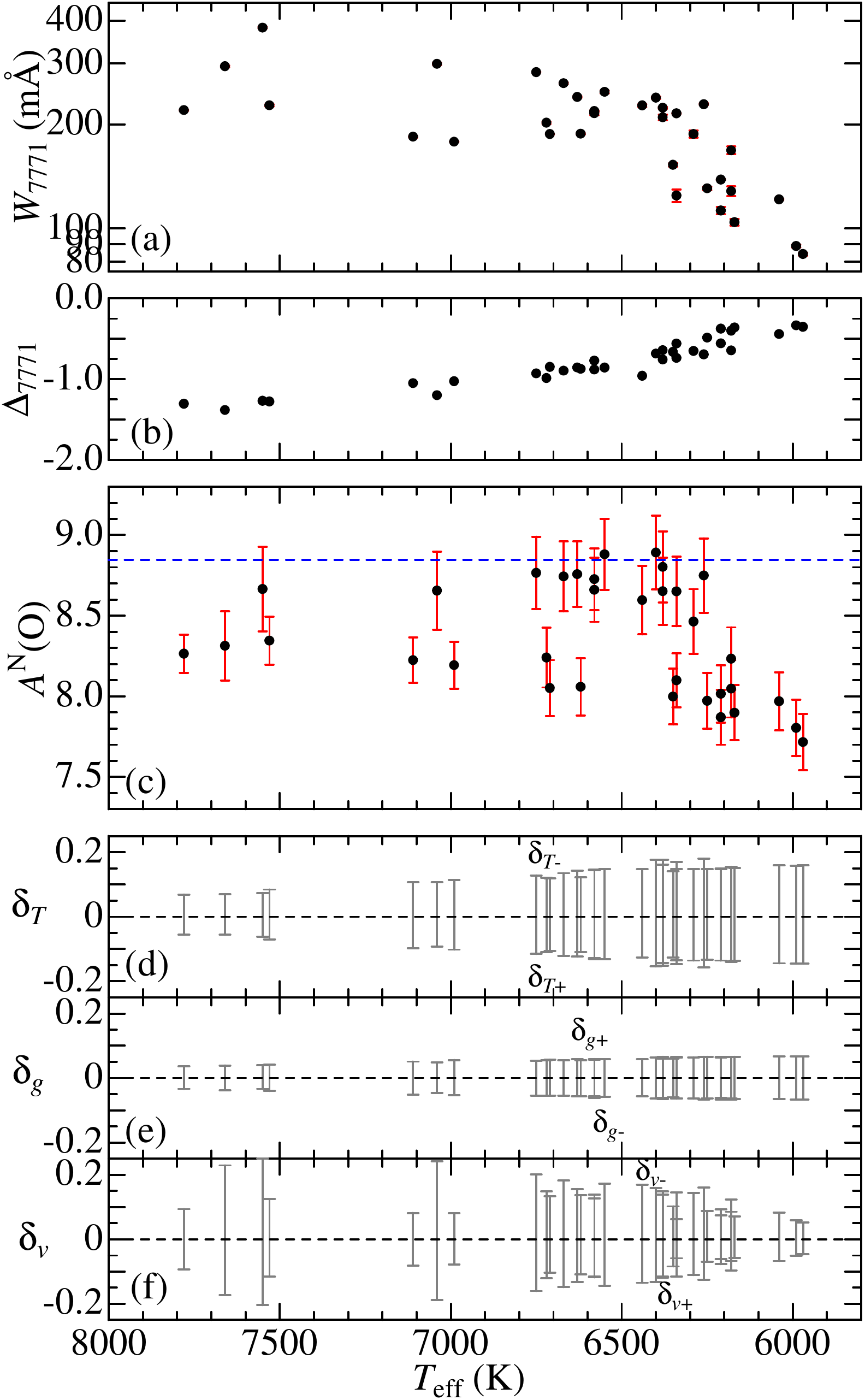}
\caption{
Oxygen abundance and O~{\sc i}~7771-related quantities 
plotted against $T_{\rm eff}$. 
The reference solar $A^{\rm N}_{\odot}$(O) value of 8.85 is indicated 
Otherwise, the same as in Fig.~7. 
}
\label{fig9}
\end{minipage}
\end{figure}

\setcounter{figure}{9}
\begin{figure}
\begin{minipage}{80mm}
\includegraphics[width=8.0cm]{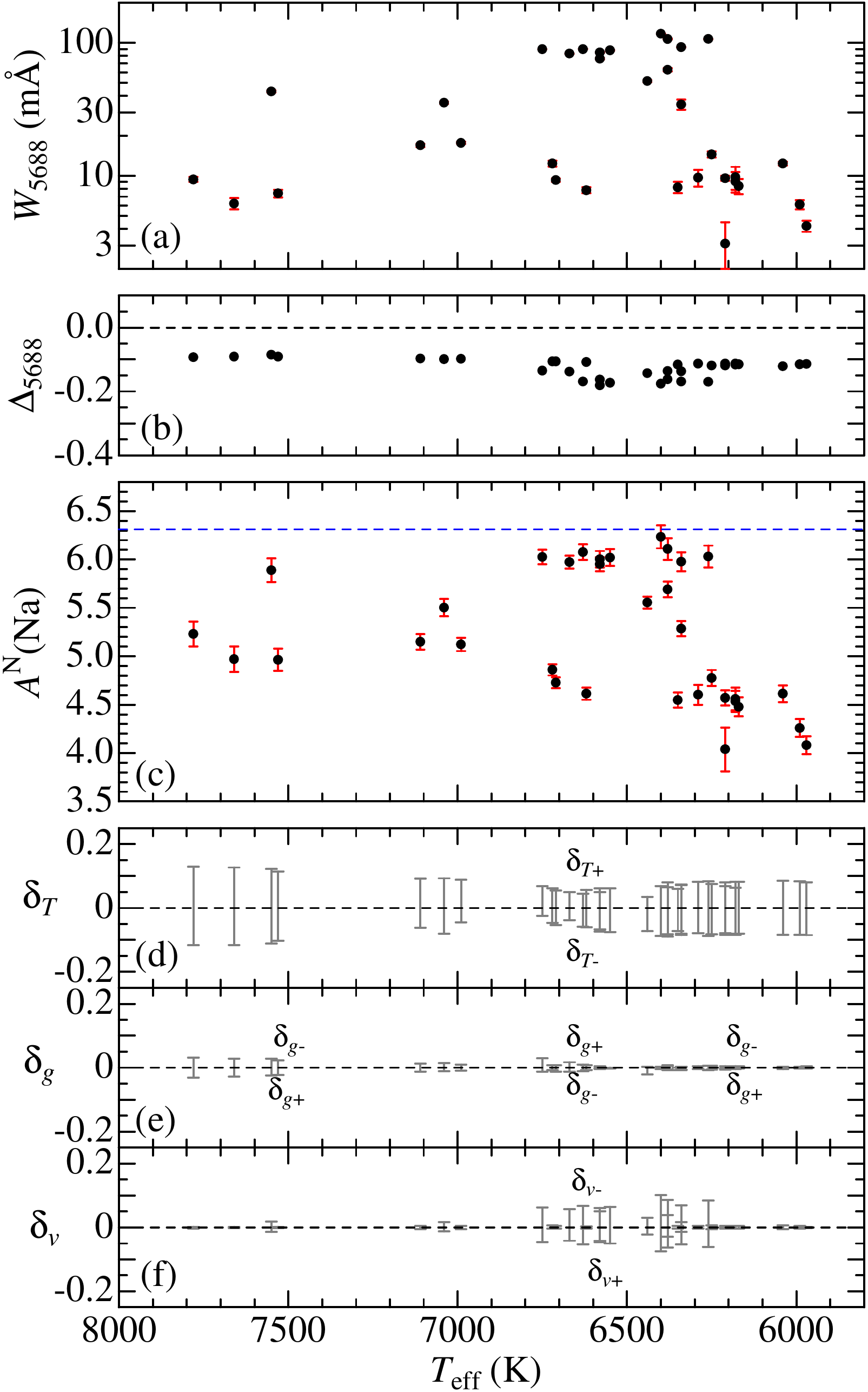}
\caption{
Sodium abundance and Na~{\sc i}~5688-related quantities 
plotted against $T_{\rm eff}$. 
The reference solar $A^{\rm N}_{\odot}$(Na) value of 6.31 is indicated 
by the horizontal dashed line in panel (c).
Otherwise, the same as in Fig.~7. 
}
\label{fig10}
\end{minipage}
\end{figure}

\subsection{Error estimation}

In order to evaluate abundance errors caused by uncertainties
in atmospheric parameters, we estimated six kinds of abundance variations
($\delta_{T+}$, $\delta_{T-}$, $\delta_{g+}$, $\delta_{g-}$, 
$\delta_{v+}$, and $\delta_{v-}$) for $A^{\rm N}$ by repeating the 
analysis on the $W$ values while 
perturbing the standard atmospheric parameters interchangeably by 
$\pm 200$~K in $T_{\rm eff}$, $\pm 0.2$~dex in $\log g$, 
and $\pm 20\%$ in $v_{\rm t}$ (which are tentatively assigned as typical 
ambiguities of the parameters according to the results in Paper~I). 
These $\delta_{T\pm}$, $\delta_{g\pm}$, and $\delta_{v\pm}$ are plotted
against $T_{\rm eff}$ in panels (d), (e), and (f) of Fig.~7--10.
It can be seen from these figures that $\delta_{T\pm}$ is generally more 
important ($\la$~0.1--0.2~dex) than $\delta_{g\pm}$ (reflecting the high-excitation 
nature of the adopted lines), while $\delta_{v\pm}$ is insignificant 
(mainly due to the weakness of lines) except for the case of O~{\sc i} 7771 line.

Further, errors due to random noises of the observed spectra were  
estimated from S/N-related uncertainties in the equivalent width ($W$) 
by invoking the relation derived by Cayrel (1988),
$\delta W \simeq 1.6 (w \delta x)^{1/2} \epsilon$,
where $\delta x$ is the pixel size (0.03--0.05~\AA), $w$ is the line width (for 
which $\lambda v_{\rm M}/c$ was tentatively assigned; 
where $\lambda$ is line wavelength, $v_{\rm M}$ is the macrobroadening velocity, 
and $c$ is the velocity of light), and $\epsilon \equiv ({\rm S/N})^{-1}$
(S/N is the signal-to-noise ratio in the neighbourhood of the line).
The resulting $\delta W$ (from a few tenths m\AA\ to several m\AA) 
are considerably smaller than $W$ and thus generally unimportant 
as seen from the size of error bars in panel (a) of Fig.~7--10,
though this error may be exceptionally significant for very weak line cases
where $\delta W$ and $W$ are comparable. 

Therefore, for the case where lines are not weak ($W \ge 20$~m\AA), the abundance 
uncertainties ($\delta A$) were evaluated by considering only the ambiguities in 
$T_{\rm eff}$, $\log g$, and $v_{\rm t}$ as 
$\delta A \equiv \delta_{Tgv} \equiv (\delta_{T}^{2} + \delta_{g}^{2} + \delta_{v}^{2})^{1/2}$,
where $\delta_{T}$, $\delta_{g}$, and $\delta_{v}$ are defined as
$\delta_{T} \equiv (|\delta_{T+}| + |\delta_{T-}|)/2$, 
$\delta_{g} \equiv (|\delta_{g+}| + |\delta_{g-}|)/2$, 
and $\delta_{v} \equiv (|\delta_{v+}| + |\delta_{v-}|)/2$,
respectively. Meanwhile, for the weak line case ($W < 20$~m\AA; where $A \propto W$), the 
contribution due to $\delta W$ was added as 
$\delta A \equiv [\delta_{Tgv}^{2} + (\delta^{'}_{W})^2]^{1/2}$,
where $\delta^{'}_{W}$ is the mean of $|\log (1 \pm \delta W/W)|$.
These $\pm\delta A$ are shown as the error bars in panel (c) of Fig.~7--10.
The values of S/N, $\delta W$, $\delta A$ for each element and for each star are also given 
in ``abunds.dat'' of the supplementary material.

For the five stars (AO~Peg, DH~Peg, DX~Del, RR~Lyr, and VY~Ser), for which several spectra 
at different phases are available, mean $\langle$[X/Fe]$\rangle$ values were calculated 
by averaging with the weight proportional to $W/\delta W$ 
(equivalent width of the relevant line divided by its error), 
while the mean $\langle$[Fe/H]$\rangle$ was evaluated
all with the same weight. These mean results are given in the last five rows in Table~1.
Generally, [X/Fe] data based on very weak lines satisfying the condition $\delta W/W > 0.5$  
were regarded as unreliable. 
Besides, quantities (e.g., [C/N] ratio or mean $\langle$[X/Fe]$\rangle$) essentially 
based on such unreliable [X/Fe] (and mean [X/Fe] values for which the standard deviation
is larger than 0.2~dex) were also judged to be trustless. 
These dubious data are parenthesised in Table~1. 

\section{Discussion}

\subsection{Observed trends of C, N, O, and Na abundances}

Let us first elucidate the characteristics of surface abundances resulting from the analysis. 
The values of [C/Fe], [O/Fe], [N/Fe], and [Na/Fe] determined for each RR~Lyr star are plotted  
against [Fe/H] in  Fig.~11a, 11b, 11c, and 11d, respectively, where the corresponding
relations of unevolved dwarfs (taken from various literature) are also shown for comparison. 
Whether and how the surface abundance of an element X (X = C, N, O, and Na) has suffered changes 
in the course of stellar evolution may be studied by comparing the trends of [X/Fe]$_{\rm RR Lyr}$ 
(circles) and [X/Fe]$_{\rm dwarfs}$ (crosses) with each other. The following tendencies
are read from these figures.
\begin{itemize}
\item
The [O/Fe] vs. [Fe/H] trend for RR~Lyr stars is essentially the same as that for dwarfs
(Fig.~11b), indicating that the surface abundance of O is almost kept without being 
appreciably affected by mixing.  
\item
While [C/Fe]$_{\rm dwarfs}$ is apt to be supersolar ($> 0$) like $\alpha$ elements, 
[C/Fe]$_{\rm RR Lyr}$ tends to be negative with a decrease in [Fe/H]
(Fig.~11a), which means an enhanced C deficiency in the surface of RR~Lyr stars towards 
lower metallicity.
\item
In contrast, while [N/Fe]$_{\rm dwarfs}$ distribute around $\sim 0$ without any clear
dependence upon metallicity, [N/Fe]$_{\rm RR Lyr}$ progressively increases with a decrease
in [Fe/H] (Fig.~11c). This suggests an enhanced enrichment of N in the surface of RR~Lyr 
stars as [Fe/H] is lowered. 
\item
Regarding Na, the behaviours of [Na/Fe]$_{\rm RR Lyr}$ and [Na/Fe]$_{\rm dwarfs}$ 
with a decrease in [Fe/H] are quite similar (slightly supersolar at first followed 
by a downturn into subsolar regime at [Fe/H]~$\la -1$) (Fig.~11d),\footnote{
Note that generally higher [Na/Fe] values were obtained in Paper~II (cf. Fig.~5 therein) 
in comparison with those of this study. This systematic discrepancy  (cf. Fig.~2e) is 
attributed to that fact that the non-LTE effect was not taken into account in the 
analysis of Paper~II.} which indicates that the primordial composition of Na is almost 
retained in the surface of RR~Lyr stars (like the case of O). 
\end{itemize} 

\subsection{Checking with theoretical predictions}

We are now able to check the mechanism of internal mixing by comparing these observed 
abundance trends with the surface abundance changes predictable from theoretical 
stellar evolution calculations.
In Fig.~12 are shown the expected surface abundance changes of C, N, O, and Na during 
the course of post-main-sequence evolution of low-mass stars (plotted against $T_{\rm eff}$) 
calculated by Lagarde et al. (2012b) for 0.85~M$_{\odot}$ (with $z= 0.0001$ and 0.002) 
and 1.0~M$_{\odot}$ (with $z= 0.0001$, 0.002, 0.004, and 0.014),
where the results for two different assumptions of internal mixing (standard 
treatment, non-standard treatment including rotational and thermohaline mixing)
are presented.
In addition, since we are primarily interested in the abundance changes at the
horizontal-branch (or red clump) phase, the phase points at the maximum $T_{\rm eff}$ 
after the He-flash are marked by filled circle at each panel, and the corresponding
values of $\log [X/X_{0}]$ are separately given in Table~4. 

\subsubsection{Carbon and nitrogen}

First of all, our attention is paid to C and N, since their surface abundances serve as 
the most important key for clarifying the physical process of evolution-induced mixing.  
It is apparent from Fig.~12 and Table~4 that a deficiency of C (Fig.~12a, 12b) 
and an enrichment of N (Fig.~12c, 12d) are generally expected. 
However, a clear difference exists between 
the two mixing cases (standard mixing, non-standard mixing); i.e., these surface 
abundance anomalies of C and N are progressively enhanced with a decrease of metallicity 
in the latter case with thermohaline mixing (while such a metallicity-dependence is not 
seen in the former case). Accordingly, the [C/N] ratio (mostly negative) is predicted to 
decrease as [Fe/H] is lowered in the latter case of non-canonical mixing, which is just
Lagarde et al. (2019) showed in their Fig.~8. 

The observed trends of [C/Fe] and [N/Fe] described in Sect.~5.1 (Figs.~11a, 11c) are 
obviously in favour of the simulation done by including the effect of thermohaline mixing.
Actually, The [C/N] vs. [Fe/H] relations derived for our sample of RR~Lyr stars are
reasonably consistent with the distributions predicted for the case of non-canonical mixing 
(taken from Fig.~8 of Lagarde et al. 2019) as shown in Fig.~11e. Accordingly, it may be 
concluded that (only the standard mixing is not sufficient but) thermohaline mixing should 
also be taken into account for evolved low-mass stars (especially at lower metallicity).

\subsubsection{Oxygen}

Regarding oxygen, stellar evolution calculations suggest that its surface abundance is hardly 
affected whichever mixing mechanism is assumed (Table~4, Figs.~12e, 12f). Therefore, the observed fact
that [O/Fe] vs. [Fe/H] trends for RR~Lyr stars and for unevolved dwarfs are practically the same
(cf. Sect.~5.1) does not contradict the theoretical expectation, though it is not informative for
elucidating the physical process of dredge-up.

\subsubsection{Sodium}

We see a problem with this element. Figs.~12g and 12h suggest that, while Na abundances
at the surface suffer little changes regardless of metallicity (at least up to the stage 
of horizontal branch) for the case of standard mixing,\footnote{
As shown in Table~4, an appreciable increase of surface Na by +0.16~dex is seen 
even in the standard mixing only for the case of $M$~=~1~M$_{\odot}$ and $z=0.002$
(its change appears to begin already at the first dredge-up phase before He ignition),
in contrast to other 
($M$, $z$) combinations where the primordial Na is retained
in the standard mixing case. Considering a possibility that this might be due to
some kind of error, we refrain from taking into account this data for the time being.  
} they are predicted to be conspicuously 
enriched at the low-metallicity regime in the case of non-canonical mixing in the sense that 
the anomaly is enhanced with a decrease in metallicity (cf. Table~4).
A comparison of these theoretical predictions with the observed results described in Sect.~5.1 
(primordial Na abundances are almost retained at the surface of RR~Lyr stars without being affected 
by evolution-induced mixing) naturally indicates that the former case of standard mixing is
preferable as long as Na is concerned. Consequently, implication from C and N (supporting
the non-canonical mixing) and that from Na (in favour of standard mixing) are inconsistent
with each other. This may suggest that Lagarde et al.'s (2012b) calculations would require 
some more appropriate improvements.   

\setcounter{figure}{10}
\begin{figure*}
\begin{minipage}{120mm}
\includegraphics[width=11.0cm]{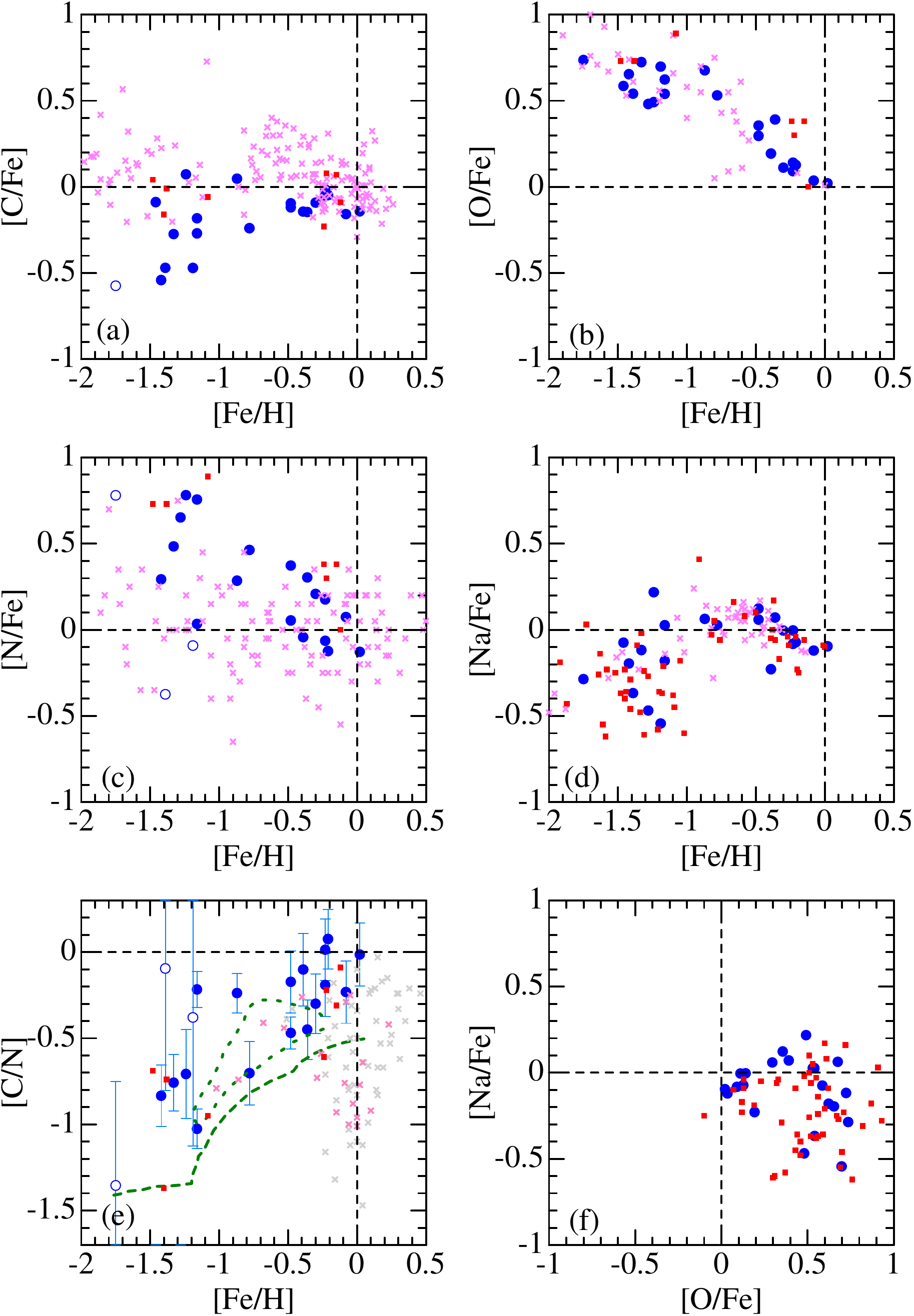}
\caption{The final results of [C/Fe], [O/Fe], [N/Fe], [Na/Fe], and [C/N]  
derived in this study for RR Lyr stars (summarised in Table~1; mean values 
are used for AO~Peg, DH~Peg, DX~Del, RR~Lyr, and VY~Ser) are plotted 
against [Fe/H] by blue circles in panels (a)--(e), respectively, 
while the relation between [Na/Fe] and [O/Fe] is depicted in panel (f). 
Note that open circles denote that these data are unreliable (i.e., parenthesised
values in Table~1) and thus should not be seriously taken.
In addition, the similar results for RR~Lyr stars derived by Andrievsky et al. (2021) 
(for C, N, O; in panels a, b, c, and e) and those by Andrievsky et al. (2018) 
(for Na; in panels d and f) are shown by small red squares.  
The error bars of [C/N] shown in panel (e) denote 
$\pm \sqrt{\delta A_{\rm C}^{2}+\delta A_{\rm N}^{2}}$ 
(cf. Sect. 4.3 for the definition of $\delta A$).  
In panels (a)--(e), the results derived for other type of metal-poor stars 
taken from various literature are also overplotted by crosses for comparison: 
(a) Takeda \& Honda's (2003) non-LTE reanalysis results of [C/Fe] for dwarfs 
based on the C~{\sc i} 7771--9 and C~{\sc i} 9061--9111 data of Tomkin et al. (1992, 1995) 
and Akerman et al. (2004). 
(b) Takeda's (2003) non-LTE reanalysis results of [O/Fe] for dwarfs (+subgiants) based on 
the O~{\sc i} 7771--5 data of Abia \& Rebolo (1989) and Mishenina et al. (2000). 
(c) Laird's (1985) [N/Fe] results for dwarfs (cf. his Table~1). 
(d) Gehren et al.'s (2006) non-LTE [Na/Fe] results for dwarfs (cf. their Table~2). 
(e) Af\c{s}ar, Sneden, \& For's (2012) [C/N] results for red horizontal-branch (RHB) 
candidates (which were calculated from CH-based [C/H] and O~{\sc i} based non-LTE 
[O/H] given in their Table~7); here, only those stars which they clearly concluded 
to be RHB are coloured in pink (otherwise shown in gray inconspicuously). 
In addition,  characteristic distribution regions of [C/N] theoretically expected 
for metal-poor giants under the influence of thermohaline mixing (roughly read 
from Fig.~8 of Lagarde et al. 2019) are shown 
by green lines (long-dashed line indicates the sharp sequence corresponding to 
red clump stars, while the.area embraced by short-dashed lines also represents 
the possible region of existence).
} 
\label{fig11}
\end{minipage}
\end{figure*}

\setcounter{figure}{11}
\begin{figure*}
\begin{minipage}{120mm}
\includegraphics[width=12.0cm]{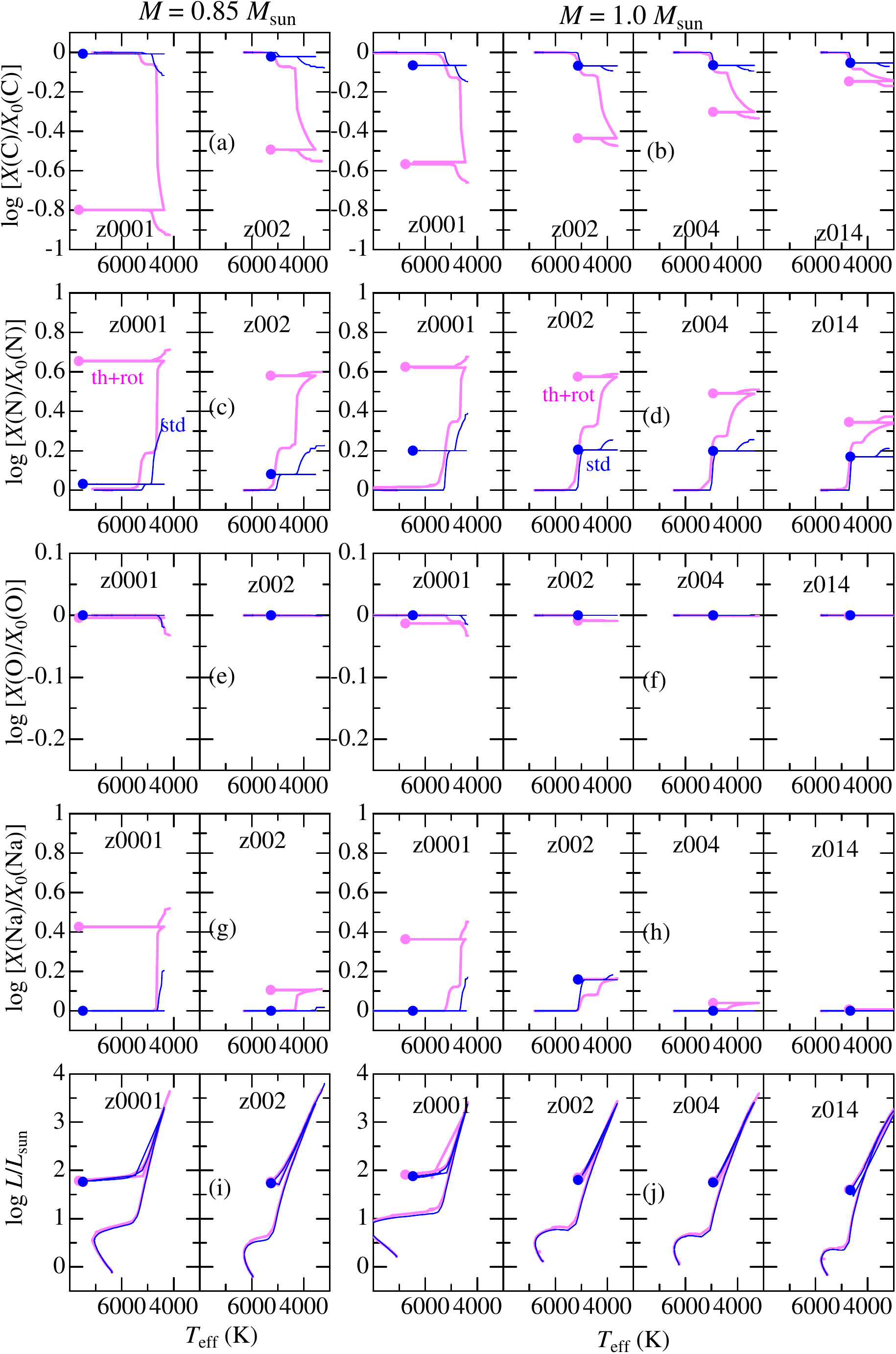}
\caption{
Run of $\log [X/X_{0}]$ (logarithmic mass fraction ratio of the 
relevant element at the surface relative to the initial value) with $T_{\rm eff}$
according to  Lagarde et al.'s (2012b) theoretical stellar evolution calculations.
Shown here are the results for $M = 0.85$~M$_{\odot}$ (left set of panels in 
two columns labeled as `z0001' and `z002' corresponding to $z$ = 0.0001 and 0.002, 
or [Fe/H] = $-2.16$ and $-0.86$) and $M = 1.0$~M$_{\odot}$ 
(right set of panels in four columns labeled as `z0001', `z002', `z004', and `z014' 
corresponding to $z$ = 0.0001, 0.002, 0.004, and 0.014 = solar metallicity, 
or [Fe/H] = $-2.16$, $-0.86$, $-0.56$, and 0.00). 
The panels in the 1st, 2nd, 3rd, and 4th row 
are the $\log [X/X_{0}]$ vs. $T_{\rm eff}$ diagrams for
C (= $^{12}$C+$^{13}$C+$^{14}$C $\simeq ^{12}$C), N (=$^{14}$N) O (= 
$^{16}$O+$^{17}$O+$^{18}$O $\simeq ^{16}$O), and Na (=$^{23}$Na), respectively,
while those in the bottom row show the corresponding $\log L$ vs. $T_{\rm eff}$
relations (theoretical Hertzsprung-Russell diagrams).
Here, the data are restricted only to those after the zero-age main sequence 
(while neglecting those of the pre-main sequence stage).
Note that two kinds of curves are shown corresponding to different treatments 
of internal mixing; i.e., standard treatment (blue lines) and non-canonical 
treatment including rotational and thermohaline mixing (pink lines).
The bullet points (filled circles) marked at each panel indicate the horizontal-branch 
(or red-clump) phase closely related to RR~Lyr stars, which is defined as the point of 
maximum $T_{\rm eff}$ after the He ignition (the corresponding $\log [X/X_{0}]$ data 
are separately presented in Table~4). 
}
\label{fig12}
\end{minipage}
\end{figure*}

\setcounter{table}{3}
\begin{table*}
\begin{minipage}{180mm}
\small
\caption{Theoretically simulated surface abundance changes of C, N, O, and Na at the horizontal branch.}
\begin{center}
\begin{tabular}{cccccc}\hline\hline
$M$ & $z$ & $\log[X/X_{0}]_{\rm C}$ & $\log[X/X_{0}]_{\rm N}$ & $\log[X/X_{0}]_{\rm O}$ & $\log[X/X_{0}]_{\rm Na}$ \\
 (M$_{\odot}$)    &        & (dex) & (dex) & (dex) & (dex) \\  
\hline
 \multicolumn{6}{c}{[standard mixing]}\\
  0.85 & 0.0001 & $-$0.0074 & +0.0315 & 0.0000 & 0.0000 \\
  0.85 &  0.002 & $-$0.0216 & +0.0810 & 0.0000 & 0.0000 \\ 
\hline
  1.00 & 0.0001 & $-$0.0662 & +0.2004 & $-$0.0001 & 0.0000 \\
  1.00 &  0.002 & $-$0.0686 & +0.2054 & $-$0.0001 & +0.1585 \\
  1.00 &  0.004 & $-$0.0655 & +0.1988 & $-$0.0001 & 0.0000 \\
  1.00 &  0.014 & $-$0.0532 & +0.1697 & 0.0000 & 0.0000 \\
\hline
 \multicolumn{6}{c}{[non-canonical mixing]}\\
  0.85 & 0.0001 & $-$0.7986 & +0.6548 & $-$0.0041 & +0.4264 \\
  0.85 &  0.002 & $-$0.4932 & +0.5804 & $-$0.0009 & +0.1058 \\
\hline
  1.00 & 0.0001 & $-$0.5667 & +0.6250 & $-$0.0132 & +0.3629 \\
  1.00 &  0.002 & $-$0.4360 & +0.5752 & $-$0.0089 & +0.1601 \\
  1.00 &  0.004 & $-$0.3023 & +0.4907 & $-$0.0013 & +0.0394 \\
  1.00 &  0.014 & $-$0.1474 & +0.3452 & $-$0.0005 & +0.0064 \\
\hline
\end{tabular}
\end{center}
Presented are the $\log [X/X_{0}]$ values (logarithmic ratio of 
the surface abundance relative to the primordial abundance) at 
the horizontal-branch (or red-clump) phase of maximum $T_{\rm eff}$ 
after the He ignition (marked by the filled circles in Fig.~12),
taken from the simulations of Lagarde et al. (2012b). 
The first six rows and the last six rows correspond to the cases 
of standard and non-standard mixing, respectively.
See also the caption of Fig.~12 for more details. 
\end{minipage}
\end{table*}

\subsection{Comparison with previous studies}

Finally, some comments may be in order regarding the results of past related investigations  
in comparison with the consequences of this study.

Regarding Andrievsky et al.'s (2021) study (similar to the present one) on the CNO abundances 
of RR~Lyr stars, 2 of their 9 sample stars are in common: DH~Peg (considerably metal-poor) and DX~Del 
(near-solar metallicity). Their results of ([Fe/H], [C/Fe], [N/Fe], [O/Fe]) are ($-1.40$, $-0.16$, 
$+1.21$, $+0.68$) for DH~Peg and  ($-0.12$, $-0.09$, $+0.00$, $+0.30$) for DX~Del.
Comparing these with our values given in Table~1, ($-1.16$, $-0.28$, $+0.76$, $+0.54$) for DH~Peg 
and ($-0.23$, $-0.04$, $-0.09$, $+0.10$) for DX~Del, both are reasonably consistent in most cases
(differences are within $\la$~0.1--0.2~dex). Only [N/Fe] for DH~Peg is the exceptional case, where 
an appreciable discrepancy ($\sim$~0.4--0.5~dex) is seen.  An inspection on their non-LTE corrections
for the N~{\sc i} 7468 line ($-0.18$ for DH~Peg and $-0.15$ for DX~Del) in comparison with
our $\langle \Delta_{7468}\rangle$ values ($-0.43$ for DH~Peg and $-0.21$ for DX~Del; 
$\langle \Delta_{7468}\rangle$ is the $W_{7468}$-weighted mean of $\Delta_{7468}$) suggests that 
the discrepancy in [N/Fe] of DH~Peg may (at least partially) be attributed to the difference in 
the applied non-LTE correction.\footnote{Andrievsky et al. (2021) reanalysed 
Takeda \& Sadakane's (1997) equivalent width data of CNO lines for the metal-poor ([Fe/H]~$\sim -1.5$) 
horizontal-branch star HD~161817. As seen from Tables~5 and 6 of their paper, their $|\Delta|$ values 
for the N~{\sc i} 8683/8686/8706 lines are generally smaller than those of Takeda \& Sadakane (1997), 
just like the case for $\Delta_{7468}$ of DH~Peg described here. Besides, they reported that their 
LTE abundances are also in disagreement. Since N abundance determinations from these N~{\sc i} 
8683/8686/8706 lines are severely affected by the strong wings of Paschen lines, the way how 
the background opacity of these overlapping H~{\sc i} lines may have been different between 
Takeda \& Sadakane (1997) and Andrievsky et al. (2021).} 

In any event, despite the existence of such a partial inconsistency, the consequences of Andrievsky et al. 
(2021) and this investigation are generally in accord with each other.  
The [Fe/H]-dependent trends of [C/Fe], [N/Fe], and [O/Fe] shown in Fig.~6 of 
their paper are consistent with our results (cf.  Fig.~11a, 11c, and 11b). 
Accordingly, the resulting [C/N] vs. [Fe/H] relation for RR~Lyr stars (cf. 
their Fig.~8) is quite similar to that obtained here (cf. Fig.~11e). 
In Fig.~11e are also overplotted the [C/N] ratios derived by Af\c{s}ar, Sneden, \& For (2012) 
for 76 stars (many are G-type giants and subgiants), where we can see that their [C/N] results 
for red horizontal-branch (RHB) stars are consistent with the trend revealed by RR~Lyr stars 
at the meal-poor regime ([Fe/H]~$\la -0.5$). Accordingly, all these results support the decreasing 
tendency of [C/N] ratios towards lower metallicity  for horizontal-branch stars (at the post 
red-giant phase), indicating the importance of thermohaline mixing in low-mass giants.
 
Turning to Na, we may invoke the work of Andrievsky et al. (2018), who studied 
the O and Na abundances of 20 field RR~Lyr stars. 
The [Na/Fe] vs. [Fe/H] relation derived by them (cf. Fig.~8 therein) is in 
satisfactory agreement with ours as shown in Fig.~11d. The mutual consistency is 
also confirmed for the correlation between [Na/Fe] and [O/Fe] (Fig.~11f).
This supports our argument that surface Na abundances 
of RR~Lyr stars have suffered little changes during the course of past evolution,
which suggested a necessity of further theoretical improvement because of the 
conflicting implications between C/N and Na (cf. Sect.~5.2.3) 

\section{Summary and conclusion}

According to recent theoretical simulations on how the nuclear-processed materials in the 
stellar interior are salvaged to the surface in the course of evolution, not only the 
standard mixing (due to deepening of convective zone) but also the thermohaline 
mixing can play a significant role for the case of low-mass stars ($M \la$~1~M$_{\odot}$)
in the sense that its importance increases with a decrease of metallicity.
This can be tested by examining the surface abundances of key elements (such as C, N, O, 
and Na; more or less affected by mixing of H-burning products) determined for evolved giant 
stars in a wide metallicity range.  

In order to check this theoretical prediction, photospheric abundances of C, N, O, and Na 
were spectroscopically determined for 22 RR~Lyr stars (low-mass horizontal-branch stars 
after He-ignition currently in the Cepheid-instability strip) based on their 34 snap-shot 
spectra obtained by High-Dispersion Spectrograph of the Subaru Telescope.

The atmospheric parameters ($T_{\rm eff}$, $\log g$, $v_{\rm t}$, and [Fe/H]) for each star 
were determined from the spectrum itself based on the equivalent widths of Fe~{\sc i}
and Fe~{\sc ii} lines, and the resulting metallicities of the sample stars are in the range 
of $-1.8 \la$~[Fe/H]~$\la 0.0$.

The C, N, O, and Na abundances were determined by applying the synthetic 
spectrum-fitting technique to the regions comprising C~{\sc i} 5380, N~{\sc i} 7468, 
O~{\sc i} 7771--5, and Na~{\sc i} 5682/5688 lines, while taking the non-LTE effect 
adequately into consideration. 
 
Comparing the resulting abundances ([C/Fe], [N/Fe], [O/Fe], and [Na/Fe]) of RR~Lyr 
stars with those of unevolved metal-poor dwarfs at the same metallicity, we were able 
to check whether and how the surface abundance of each element has suffered a change 
due to evolution-induced mixing, by which information on the physical process of 
mixing may be gained in comparison with theoretical simulations.

Regarding C and N, a characteristic trend was confirmed that C is deficient while 
N is enriched, and the extent of these peculiarities tending to increase with a decrease 
in [Fe/H]. Accordingly, the [C/N] abundance ratio progressively decreases with a lowering 
of metallicity from $\sim 0$ ([Fe/H]~$\sim 0$) to $\sim -1$ ([Fe/H]~$\sim -1.5$). 
This lends support to the theoretical prediction for the case of non-canonical mixing 
(including thermohaline mixing).  

In contrast, the observed abundances of O as well as Na in RR~Lyr stars were found 
to be almost the same as those of dwarfs, which means that the surface compositions 
of these two elements are almost retained without being affected by dredge-up of 
nuclear-processed products.

The inertness of O is reasonably understandable, which is anyhow theoretically 
expected for whichever case of standard or non-canonical mixing. 
Meanwhile, according to currently available simulations, surface Na abundance is 
predicted to be progressively more enriched towards lower metallicity for the case 
of non-canonical mixing, while it hardly suffers changes for the case of standard mixing.
Therefore, the observed behaviour of Na is more consistent with the case of 
standard mixing (i.e., without thermohaline mixing).  

Consequently, since implications from C/N (supporting non-canonical mixing) and 
Na (in favour of standard mixing) are rather contradictory, this presumably means
that the current theory is not complete and still has room for further improvement.
The results obtained in this investigation may hopefully serve as observational 
constraints towards further progress on the theoretical side.

\section*{Acknowledgements}

This investigation is based on the data collected at Subaru Telescope, 
which is operated by the National Astronomical Observatory of Japan.
This research has made use of the SIMBAD database, operated by CDS, 
Strasbourg, France. 

\section*{Data availability}

The basic data and results underlying this article are presented as 
the online supplementary material. The original observational data
are in the public domain and available at https://smoka.nao.ac.jp/index.jsp 
(SMOKA Science Archive site).

\section*{Supporting information}

Additional Supporting Information may be found in the supplementary materials.
\begin{itemize}
\item
{\bf ReadMe} 
\item
{\bf felines.dat} 
\item
{\bf abunds.dat} 
\end{itemize}


\begin{thebibliography}{}
\bibitem[]{} 
  Abia C., Rebolo R., 1989, ApJ, 347, 186
\bibitem[]{}
  Af\c{s}ar M., Sneden C., For B.-Q., 2012, AJ, 144, 20
\bibitem[]{}
  Akerman C.~J., Carigi L., Nissen P.~E., Pettini M., Asplund M.,
  2004, A\&A, 414, 931
\bibitem[]{}
  Allende Prieto C., Barklem P.~S., Lambert D.~L., Cunha K., 2004,
  A\&A, 420, 183
\bibitem[]{}
  Andrievsky S. et al., 2018, PASP, 130, 024201
\bibitem[]{}
  Andrievsky S.~M., Korotin S.~A., Kovtyukh V.~V., Khrapaty S.~V., Rudyak Y., 2021, 
  Astron. Nachr., 342, 887
\bibitem[]{}
  Cayrel R. 1988, in The Impact of Very High S/N Spectroscopy on Stellar Physics,
  Proc. IAU Symp. 132, eds. G. Cayrel de Strobel \& M. Spite (IAU), p. 345
\bibitem[]{}
  Charbonnel C., Lagarde N., 2010, A\&A, 522, A10
\bibitem[]{}
  Gehren T., Shi J.~R., Zhang H.~W., Zhao G., Korn A.~J., 2006, A\&A, 451, 1065
\bibitem[]{}
  Kurucz R.~L., 1993, Kurucz CD-ROM, No. 13, ATLAS9 Stellar Atmosphere
  Program and 2 km/s Grid (Cambridge, MA: Harvard-Smithsonian Center
  for Astrophysics)
\bibitem[]{}
  Kurucz R.~L., Bell B., 1995, Kurucz CD-ROM, No. 23, Atomic Line List
  (Cambridge, MA: Harvard-Smithsonian Center for Astrophysics)
\bibitem[]{}
  Kurucz R.~L., Furenlid I., Brault J., Testerman L., 1984,  
  Solar Flux Atlas from 296 to 1300 nm
  (Sunspot, New Mexico: National Solar Observatory)
\bibitem[]{}
  Lagarde N., Romano D., Charbonnel C., Tosi M., Chiappini C.,
  Matteucci F., 2012a, A\&A, 542, A62
\bibitem[]{}
  Lagarde N., Decressin T., Charbonnel C., Eggenberger P., Ekstr\"{o}m S.,
  Palacios A., 2012b, A\&A, 543, A108
\bibitem[]{}
  Lagarde N. et al., 2019, A\&A, 621, A24
\bibitem[]{}
  Laird J.~B., 1985, ApJ, 289, 556
\bibitem[]{}
  Liu S., Zhao G., Chen Y.-Q., Takeda Y., Honda S., 2013, 
  Research in Astron. Astrophys., 13, 1307  (Paper~II)
\bibitem[]{} 
  Mishenina T.~V., Korotin S.~A., Klochkova V.~G., Panchuk, V.~E., 
  2000, A\&A, 353, 978
\bibitem[]{}
  Takeda Y., 1995, PASJ, 47, 287
\bibitem[]{}
  Takeda Y., 2003, A\&A, 402, 343
\bibitem[]{}
  Takeda Y., et al., 2005a, PASJ, 57, 13
\bibitem[]{}
  Takeda Y., Honda S., 2005, PASJ, 57, 65
\bibitem[]{}
  Takeda Y., Honda S., Aoki W., Takada-Hidai M., Zhao G., Chen Y.-Q., 
  Shi J.-R., 2006, PASJ, 58, 389 (Paper~I)
\bibitem[]{}
  Takeda Y., Kawanomoto S., Sadakane K., 1998, PASJ, 50, 97
\bibitem[]{}
  Takeda Y., Ohkubo M., Sato B., Kambe E., Sadakane K., 2005b, PASJ, 57, 27
\bibitem[]{}
  Takeda Y., Sadakane K., 1997, PASJ, 49, 571
\bibitem[]{}
  Takeda Y., Zhao G., Takada-Hidai M., Chen Y.-Q., Saito Y.-J.,
  Zhang H.-W., 2003, ChJAA, 3, 316
\bibitem[]{}
  Tomkin J., Lemke M., Lambert D.~L., Sneden C., 1992, AJ, 104, 1568
\bibitem[]{}
  Tomkin J., Woolf V.~M., Lambert D.~L., Lemke M., 1995, AJ, 109, 2204
\end{thebibliography}
\end{document}